\documentclass[preprint]{aastex}
\usepackage{graphicx}
\newcommand{\mjup}{M$_{Jup}$}
\slugcomment{To be submitted to \apj .}

\shorttitle{$L'$ Exoplanet Survey of Sunlike Stars: Obs}
\shortauthors{Heinze et al.}

\begin{document}

\title{Constraints on Long-Period Planets from an $L'$ and
$M$ band Survey of Nearby Sun-Like Stars: Observations\altaffilmark{1}}

\author{A. N. Heinze}
\affil{Steward Observatory, University of Arizona, 933 N Cherry Avenue, Tucson, AZ 85721}
\email{ariheinze@hotmail.com}
\author{Philip M. Hinz}
\affil{Steward Observatory, University of Arizona, 933 N Cherry Avenue,
  Tucson, AZ 85721}
\email{phinz@as.arizona.edu}
\author{Suresh Sivanandam}
\affil{Steward Observatory, University of Arizona, 933 N Cherry Avenue,
  Tucson, AZ 85721}
\email{suresh@as.arizona.edu}
\author{Matthew Kenworthy}
\affil{Steward Observatory, University of Arizona, 933 N Cherry Avenue,
  Tucson, AZ 85721}
\email{mmeyer@phys.ethz.ch}
\author{Michael Meyer}
\affil{Department of Physics, Swiss
Federal Institute of Technology (ETH-Zurich), ETH H\'{o}nggerberg,
CH-8093 Zurich, Switzerland}
\email{mmeyer@as.arizona.edu}
\author{Douglas Miller}
\affil{Steward Observatory, University of Arizona, 933 N Cherry Avenue,
  Tucson, AZ 85721}
\email{dlmiller@as.arizona.edu}

\altaffiltext{1}{Observations reported here were obtained at the MMT
Observatory, a joint facility of the University of Arizona and the
Smithsonian Institution.}

\begin{abstract}

We present the observational results of an $L'$ and
$M$ band Adaptive Optics (AO) imaging survey of 54
nearby, sunlike stars for extrasolar planets, carried 
out using the Clio camera on the MMT.  
We have concentrated more strongly than
all other planet imaging surveys to date on very nearby F, G, and K stars,
prioritizing stellar proximity higher than youth. Ours is
also the first survey to include extensive observations in the $M$
band, which supplement the primary $L'$ observations.  Models predict
much better planet/star flux ratios at the $L'$ and
$M$ bands than at more commonly used shorter wavelengths (i.e. the $H$ band).  
We have carried out extensive blind simulations with fake
planets inserted into the raw data to verify our sensitivity,
and to establish a definitive relationship between source
significance in $\sigma$ and survey completeness.  We find
97\% confident-detection completeness for 10$\sigma$ sources,
but only 46\% for 7$\sigma$ sources -- raising concerns about the
standard procedure of assuming high completeness at 5$\sigma$,
and demonstrating that blind sensitivity tests to establish
the significance-completeness relation are an important 
analysis step for all planet-imaging
surveys.  We discovered a previously unknown $\sim0.15$M$_{\sun}$  
stellar companion to the F9 star GJ 3876, at a 
projected separation of about 80 AU.  Twelve additional
candidate faint companions are detected around other stars.  Of these,
eleven are confirmed to be background stars, and one is a previously
known brown dwarf.  We obtained sensitivity to planetary-mass
objects around almost all of our target stars, with sensitivity
to objects below 3 \mjup~in the best cases.  Constraints on
planet populations based on this null result are presented in
our Modeling Results paper, \citet{modeling}.

\end{abstract}


\section{Introduction}

Nearly 400 extrasolar planets have now been discovered
using the radial velocity (RV) method.  RV surveys currently
have good statistical completeness only for planets with 
periods of less than ten years \citep{cumming,butlercat}, 
due to the limited temporal baseline of the observations, 
and the need to observe for a complete orbital period to 
confirm the properties of a planet with confidence. 
The masses of discovered planets range from
just a few Earth masses \citep{hotNep} up to around 20 Jupiter
masses (\mjup).  We note that a 20 \mjup~object would be considered
by many to be a brown dwarf rather than a planet, but that there is 
no broad consensus on how to define the upper mass limit for
planets.  For a good overview of RV planets to date, see
\citet{butlercat} or \url{http://exoplanet.eu/catalog-RV.php}.

The large number of RV planets has enabled several good
statistical analyses of planet populations \citep{carnp,
butlercat,cumming}.  However, these apply only to the
short-period planets accessible to RV surveys.  We cannot 
obtain a good understanding of planets in general 
without information on long period extrasolar planets;
nor can we see how our own solar system fits into
the big picture of planet formation in the galaxy
without a good census of planets in Jupiter- and
Saturn-like long-period orbits around other stars.

Several methods (transit detection,
RV variations, astrometry, and direct
imaging) have yielded repeatable
detections of extrasolar planets so far.  While RV
and astrometric surveys may eventually deliver important
information about long-period extrasolar planets,
direct imaging is the only method that allows us to characterize
them on a timescale of months rather than years or decades.

Direct imaging of extrasolar planets is technologically
possible at present only in the infrared, based on
the planets' own thermal luminosity, not on reflected
starlight.  The enabling technology is adaptive optics (AO),
which allows 6-10m ground-based telescopes to obtain diffraction
limited IR images several times sharper than those from
HST, despite Earth's turbulent atmosphere.  Theoretical
models of giant planets indicate that
such telescopes should be capable of detecting self-luminous
giant planets in large orbits around young, nearby stars.
The stars should be young because the glow of
giant planets comes from gravitational potential
energy converted to heat in their formation and
subsequent contraction: lacking any internal fusion,
they cool and become fainter as they age.

Several groups have published the results of AO imaging
surveys for extrasolar planets around F, G, K, or M stars
in the last five years (see for example \citet{masciadri,kasper,biller1,GDPS,chauvin}).  
Of these, most have used wavelengths in the 1.5-2.2 $\mu$m
range, corresponding to the astronomical $H$ and $K_S$
filters \citep{masciadri,biller1,GDPS,chauvin}.  They have targeted
mainly very young stars.  Because young stars are rare, the median
distance to stars in each of these surveys has been
more than 20 pc.

In contrast to those above, our survey concentrates on 
very nearby F, G, and K stars, with proximity prioritized more than 
youth in the sample selection.  The median distance to our
survey targets is only 11.2 pc.  Ours is also the first survey 
to include extensive observations in the $M$
band, and only the second to search solar-type stars in the $L'$
band (the first was \citet{kasper}).  The distinctive focus on older, 
very nearby stars for a survey using longer wavelengths is 
natural: longer wavelengths are optimal for detecting objects
with very red IR colors -- that is, low temperature planets.
These are most likely to be found in older systems, since
planets cool and redden with age \citep{bar,bur}.  However
old, low-temperature planets also have low luminosities,
rendering them undetectable around all but the nearest stars.  

In Section \ref{sec:samp} we describe the criteria
used in choosing our sample, and present the characteristics
of our stars.  In Section \ref{sec:Observe}, we briefly
describe our instrument, our observing strategy, and our
image processing pipeline.  In Section \ref{sec:sensanal} we
detail our sensitivity estimation methods, and show how
we characterized them using blind tests in which simulated
planets were inserted into our raw data -- a practice that should
be standard for planet imaging surveys.  In Section \ref{sec:faintreal} 
we give astrometric and photometric data for all the faint companions
detected in our survey, as well as precise astrometry of the
bright known binary stars in our sample.  We present our 
conclusions in Section \ref{sec:concl}.  Constraints on
planet populations based on our survey null result are
presented in \citet{modeling}.

\section{The Survey Sample} \label{sec:samp}
The goal of our sample selection was to pick the nearest stars
around which we could detect planets of 10 \mjup~or below.
This practically meant that very nearby stars were
potential targets up to ages of several Gyr, while at larger distances
we would consider only fairly young stars.  We set out initially
to investigate only FGK stars within 25pc of the sun, in order
to make our sample comparable in spectral type to the samples
of the RV surveys, and to focus on the nearest stars, at which
the $L'$ and $M$ bands are most useful relative to shorter
wavelengths.  In the end we included a few M stars and a few 
stars slightly beyond 25pc, because these stars were very interesting 
and we had exhausted most of the
observable stars that lay within our more strict criteria.
The stars of our sample are presented in Tables \ref{tab:samp01}
and \ref{tab:samp02}.

Our survey focuses on markedly more nearby stars than all other
surveys published to date.  For example, the median
distance to stars in the \citet{masciadri} survey is 21.2 pc.
For the \citet{kasper} survey the median distance is 37 pc,
for \citet{biller1} it is 24.7 pc, and for \citet{GDPS} it is 21.7 pc.  
Our median distance is 11.2 pc.

\begin{deluxetable}{lccccc}
\tablewidth{0pc}
\tablecolumns{6}
\tablecaption{Age, Distance, and Spectral Type of Survey Targets \label{tab:samp01}}
\tablehead{ & \colhead{Age 1} & \colhead{Age 2} & \colhead{Adopted} &
  \colhead{Dist.} & \colhead{Spectral} \\
\colhead{Star} & \colhead{(Gyr)} & \colhead{(Gyr)} & \colhead{Age (Gyr)} &
\colhead{(pc)} & \colhead{Type}}
\startdata
GJ 5 & 0.11\tablenotemark{a} & 0.2\tablenotemark{b} & 0.155 & 14.25 & K0Ve \\
HD 1405 & 0.1-0.2\tablenotemark{c} & 0.03-0.08\tablenotemark{d} &
0.1\phantom{00} & 30\phantom{.00} & K2V \\
$\tau$ Ceti & \nodata & \nodata & 5\phantom{.000} & 3.50 & G8Vp \\
GJ 117 & 0.1\tablenotemark{c} & 0.03\tablenotemark{a} & 0.1\phantom{00} & 8.31
& K2V \\
$\epsilon$ Eri & 0.56\tablenotemark{a} & \nodata & 0.56\phantom{0} & 3.27 & K2V \\
GJ 159 & 0.03-0.01\tablenotemark{e} & \nodata & 0.1\phantom{00} & 18.12 & F6V \\
GJ 166 B & \nodata & \nodata & 2\phantom{.000} & 4.83 & DA \\
GJ 166 C & \nodata & \nodata & 2\phantom{.000} & 4.83 & dM4.5e \\
HD 29391 & 0.01-0.03\tablenotemark{f} & \nodata & 0.1\phantom{00} & 14.71 & F0V \\
GJ 211 & 0.52\tablenotemark{a} & \nodata & 0.52\phantom{0} & 12.09 & K1Ve \\
GJ 216 A & 0.4-0.6\tablenotemark{g} & \nodata & 0.44\phantom{0} & 8.01 & F6V \\
BD+20 1790 & 0.06-0.3\tablenotemark{e} & \nodata & 0.18\phantom{0} & 24\phantom{.00} & K3 \\
GJ 278 C & 0.1-0.3\tablenotemark{h} & \nodata & 0.2\phantom{00} & 14.64 & M0.5Ve \\
GJ 282 A & 0.49\tablenotemark{a} & 0.4-0.6\tablenotemark{g} & 0.5\phantom{00} & 13.46 & K2Ve \\
GJ 311 & 0.1\tablenotemark{c} & 0.1-0.3\tablenotemark{e} & 0.24\phantom{0} & 13.85 & G1V \\
HD 77407 A & 0.05\tablenotemark{i} & \nodata & 0.1\phantom{00} & 30.08 & G0V \\
HD 77407 B & 0.05\tablenotemark{i} & \nodata & 0.1\phantom{00} & 30.08 & M2V \\
HD 78141 & 0.1-0.2\tablenotemark{c} & \nodata & 0.15\phantom{0} & 21.4\phantom{0} & K0 \\
GJ 349 & 0.37\tablenotemark{a} & \nodata & 0.37\phantom{0} & 11.29 & K3Ve \\
GJ 355 & 0.1\tablenotemark{c} & 0.05-0.15\tablenotemark{j} & 0.1\phantom{00} & 19.23 & K0 \\
GJ 354.1 A & 0.1\tablenotemark{c} & 0.02-0.15\tablenotemark{j} & 0.1\phantom{00} & 18.87 & dG9 \\
GJ 380 & \nodata & \nodata & 2\phantom{.000} & 4.69 & K2Ve \\
GJ 410 & 0.2-0.6\tablenotemark{g} & \nodata & 0.37\phantom{0} & 11\phantom{.00} & dM2e \\
HD 96064 A & 0.1-0.2\tablenotemark{c} & \nodata & 0.15\phantom{0} & 24.63 & G5V \\
HD 96064 B & 0.1-0.2\tablenotemark{c} & \nodata & 0.15\phantom{0} & 24.63 & M3V \\
GJ 450 & $<$1.0\tablenotemark{k} & \nodata & 1\phantom{.000} & 8.1\phantom{0} & M1Ve \\
BD+60 1417 & 0.1-0.2\tablenotemark{c} & \nodata & 0.15\phantom{0} & 17.7\phantom{0} & K0 \\
HD 113449 & 0.1-0.2\tablenotemark{c} & \nodata & 0.15\phantom{0} & 22.1\phantom{0} & G5V \\
GJ 505 A & 0.79\tablenotemark{a} & \nodata & 0.79\phantom{0} & 11.9\phantom{0} & K2V \\
GJ 505 B & 0.79\tablenotemark{a} & \nodata & 0.79\phantom{0} & 11.9\phantom{0} & M0.5V \\
GJ 519 & 0.2-0.6\tablenotemark{g} & \nodata & 0.37\phantom{0} & 9.81 & dM1 \\
GJ 3860 & 0.28\tablenotemark{a} & 0.2-0.6\tablenotemark{g} & 0.28\phantom{0} & 14.93 & K0 \\
GJ 564 & 0.1-0.2\tablenotemark{c} & \nodata & 0.15\phantom{0} & 17.94 & G2V \\
GJ 3876 & \nodata & \nodata & 2\phantom{.000} & 43.3\phantom{0} & F9IV \\
$\xi$ Boo A & 0.43\tablenotemark{a} & 0.1\tablenotemark{c} & 0.29\phantom{0} & 6.71 & G8V \\
$\xi$ Boo B & 0.15\tablenotemark{a} & \nodata & 0.29\phantom{0} & 6.71 & K4V \\
HD 139813 & 0.1-0.2\tablenotemark{c} & \nodata & 0.15\phantom{0} & 21.7\phantom{0} & G5 \\
GJ 625 & 0.4-0.6\tablenotemark{g} & \nodata & 0.5\phantom{00} & 6.28 & dM2 \\
GJ 659 A & $<$1.0\tablenotemark{l} & \nodata & 1\phantom{.000} & 20.2\phantom{0} & K8V \\
GJ 659 B & $<$1.0\tablenotemark{l} & \nodata & 1\phantom{.000} & 20.2\phantom{0} & dK8 \\
GJ 684 A & 0.4-0.6\tablenotemark{g} & \nodata & 0.5\phantom{00} & 14.09 & G0V \\
GJ 684 B & 0.4-0.6\tablenotemark{g} & \nodata & 0.5\phantom{00} & 14.09 & K3V \\
GJ 702 A & \nodata & \nodata & 2\phantom{.000} & 5.03 & K0V \\
GJ 702 B & \nodata & \nodata & 2\phantom{.000} & 5.03 & K4V \\
61 Cyg A & \nodata & \nodata & 2\phantom{.000} & 3.46 & K5V \\
61 Cyg B & \nodata & \nodata & 2\phantom{.000} & 3.46 & K7V \\
BD+48 3686 & 0.1-0.2\tablenotemark{c} & \nodata & 0.15\phantom{0} & 23.6\phantom{0} & K0 \\
GJ 860 A & $<$1.0\tablenotemark{k} & \nodata & 1\phantom{.000} & 4.01 & M2V \\
GJ 860 B & $<$1.0\tablenotemark{k} & \nodata & 1\phantom{.000} & 4.01 & M6V \\
GJ 879 & 0.1-0.3\tablenotemark{h} & \nodata & 0.2\phantom{00} & 7.81 & K5Ve \\
HD 220140 A & 0.025-0.15\tablenotemark{j} & \nodata & 0.1\phantom{00} & 19.74 & G9V \\
HD 220140 B & 0.025-0.15\tablenotemark{j} & \nodata & 0.1\phantom{00} & 19.74 &
G9V \\
GJ 896 A & $<$0.3\tablenotemark{h} & \nodata & 0.3\phantom{00} & 6.58 & M3.5 \\
GJ 896 B & $<$0.3\tablenotemark{h} & \nodata & 0.3\phantom{00} & 6.58 & M4.5 \\
\enddata
\tablecomments{The adopted age, usually an average of
the referenced values, is the age we used in our
Monte Carlo simulations.  Distances are from
\citet{hip} parallaxes.  For stars for which we
did not have specific age estimates, we adopted
an age of 2 Gyr, based on dynamical considerations
setting the mean age of thin-disk stars in the solar
neighborhood near this value; see \citet{2Gyr1} and \citet{2Gyr2}.
Admittedly this is a very approximate procedure, and
2 Gyr might be younger than the average age of the specific
systems in question -- however, these systems are not
extremely important to the overall results of the survey,
accounting in all for only 6.5\% of the total planet detection
potential, according to Table 3 of \citet{modeling}.}
\tablenotetext{a}{\footnotesize{\citet{fischer}}}
\tablenotetext{b}{\footnotesize{\citet{bryden}}}
\tablenotetext{c}{\footnotesize{\citet{wichmann03a}}}
\tablenotetext{d}{\footnotesize{\citet{lopez}}}
\tablenotetext{e}{\footnotesize{Age estimate from FEPS target
list, courtesy M. Meyer.}}
\tablenotetext{f}{\footnotesize{\citet{zuckerman01}}}
\tablenotetext{g}{\footnotesize{\citet{king03}}}
\tablenotetext{h}{\footnotesize{\citet{BYN98}}}
\tablenotetext{i}{\footnotesize{\citet{wichmann03b}}}
\tablenotetext{j}{\footnotesize{\citet{montes01}}}
\tablenotetext{k}{\footnotesize{The \citet{hunsch} 
catalog reports a ROSAT detection at a
flux level that suggests an age of 1 Gyr or less.}}
\tablenotetext{l}{\footnotesize{\citet{favata}}}
\end{deluxetable}

Surveying nearby, older stars at long wavelengths is interesting
for several reasons.  First, nearby stars offer the
best chance to see planets at small physical separations,
perhaps even inward to the outer limits of RV sensitivity.
Second, planetary systems with ages up to several hundred Myr
may still be undergoing substantial dynamical evolution
due to planet-planet interactions \citep{juric,levison}.
While finding systems in the process of dynamical evolution
would be fascinating, we also need information about
systems old enough to have settled down into a mature,
stable configuration.  To probe long-period planet
populations in mature systems, surveys such as ours
that target older stars are necessary.  

Additionally, 
theoretical spectra of older planets are likely
more reliable than for younger ones, as these planets are further
from their unknown starting conditions and moving toward a well-understood,
stable configuration such as Jupiter's.  It has been suggested by \citet{faintJup}, 
in fact, that theoretical planet models such as those of \citet{bur}
may overpredict the brightness of young ($<$ 100 Myr) planets 
by orders of magnitude, while for older planets the models are more accurate.
Lastly, $L'$ surveys such as ours and that
of \citet{kasper} are an important complement to the
shorter-wavelength work of \citet{masciadri,biller1,chauvin};
and \citet{GDPS}
in that they ensure that limits on planet populations
do not depend entirely on yet-untested predictions of the
flux from extrasolar giant planets in a narrow wavelength
interval.  Until a sufficient number of extrasolar planets
have been directly imaged that their spectra are well
understood, surveys conducted at a range of different
wavelengths will increase the confidence that may be placed
in the results.

\begin{center}
\begin{deluxetable}{lcccccc}
\tablewidth{0pc}
\tablecolumns{7}
\tablecaption{Position and Magnitude of Survey Targets\label{tab:samp02}}
\tablehead{\colhead{Star} & \colhead{RA} & \colhead{DEC} & \colhead{V} & \colhead{H} & \colhead{K} & \colhead{L'}}
\startdata
GJ 5 & 00:06:36.80 & 29:01:17.40 & 6.13 & 4.69 & 4.31 & 4.25 \\
HD 1405 & 00:18:20.90 & 30:57:22.00 & 8.60 & 6.51 & 6.39 & 6.32   \\
$\tau$ Ceti & 01:44:04.10 & -15:56:14.90 & 3.50 & 1.77 & 1.70 & 1.65 \\
GJ 117 & 02:52:32.10 & -12:46:11.00 & 6.00 & 4.23 & 4.17 & 4.11 \\
$\epsilon$ Eri & 03:32:55.80 & -09:27:29.70 & 3.73 & 1.88 & 1.78 & 1.72 \\
GJ 159 & 04:02:36.70 & -00:16:08.10 & 5.38 & 4.34 & 4.18 & 4.14   \\
GJ 166 B & 04:15:21.50 & -07:39:22.30 & 9.50 & \nodata & \nodata & \nodata \\
GJ 166 C & 04:15:21.50 & -07:39:22.30 & 11.17 & 5.75 & 5.45 & 5.05 \\
HD 29391 & 04:37:36.10 & -02:28:24.80 & 5.22 & 4.77 & 4.54 & 4.51 \\
GJ 211 & 05:41:20.30 & 53:28:51.80 & 6.23 & 3.99 & 4.27 & 4.21   \\
GJ 216 A & 05:44:27.80 & -22:26:54.20 & 3.60 & 2.47 & 2.42 & 2.38 \\
BD+20 1790 & 07:23:43.60 & 20:24:58.70 & 9.93 & 7.61 & 7.51 & 7.42 \\
GJ 278 C & 07:34:37.40 & 31:52:09.80 & 9.07 & 5.42 & 5.24 & 5.05 \\
GJ 282 A & 07:39:59.30 & -03:35:51.00 & 7.20 & 5.06 & 4.89 & 4.82 \\
GJ 311 & 08:39:11.70 & 65:01:15.30 & 5.65 & 4.28 & 4.17 & 4.12 \\
HD 77407 A & 09:03:27.10 & 37:50:27.50 & 7.10 & 5.53 & 5.44 & 5.39 \\
HD 77407 B & 09:03:27.10 & 37:50:27.50 & \nodata & \nodata & \nodata & \nodata \\
HD 78141 & 09:07:18.10 & 22:52:21.60 & 7.99 & 5.92 & 5.78 & 5.72 \\
GJ 349 & 09:29:54.80 & 05:39:18.50 & 7.22 & 5.00 & 4.79 & 4.70 \\
GJ 355 & 09:32:25.60 & -11:11:04.70 & 7.80 & 5.60 & 5.45 & 5.39 \\
GJ 354.1 A & 09:32:43.80 & 26:59:18.70 & 7.01 & 5.24 & 5.12 & 5.06 \\
GJ 380 & 10:11:22.10 & 49:27:15.30 & 6.61 & 3.93 & 2.96 & 2.89   \\
GJ 410 & 11:02:38.30 & 21:58:01.70 & 9.69 & 5.90 & 5.69 & 5.46   \\
HD 96064 A & 11:04:41.50 & -04:13:15.90 & 7.64 & 5.90 & 5.80 & 5.75   \\
HD 96064 B & 11:04:41.50 & -04:13:15.90 & \nodata & \nodata & \nodata & \nodata   \\
GJ 450 & 11:51:07.30 & 35:16:19.30 & 9.78 & 5.83 & 5.61 & 5.40   \\
BD+60 1417 & 12:43:33.30 & 60:00:52.70 & 9.40 & 7.36 & 7.29 & 7.23  \\ 
HD 113449 & 13:03:49.70 & -05:09:42.50 & 7.69 & 5.67 & 5.51 & 5.46  \\
GJ 505 A & 13:16:51.10 & 17:01:01.90 & 6.52 & 4.58 & 4.38 & 4.31  \\
GJ 505 B & 13:16:51.10 & 17:01:01.90 & 9.80 & 5.98 & 5.75 & 5.43  \\
GJ 519 & 13:37:28.80 & 35:43:03.90 & 9.07 & 5.66 & 5.49 & 5.28  \\
GJ 3860 & 14:36:00.60 & 09:44:47.50 & 7.51 & 5.63 & 5.55 & 5.49 \\
GJ 564 & 14:50:15.80 & 23:54:42.60 & 5.88 & 4.47 & 4.42 & 4.37  \\
GJ 3876 & 14:50:20.40 & 82:30:43.00 & 5.64 & 4.19 & 3.92 & 3.87   \\
$\xi$ Boo A & 14:51:23.40 & 19:06:01.70 & 4.55 & 2.82 & 2.75 & 2.70 \\
$\xi$ Boo B & 14:51:23.40 & 19:06:01.70 & 6.97 & 4.45 & 4.34 & 4.24 \\
HD 139813 & 15:29:23.60 & 80:27:01.00 & 7.31 & 5.56 & 5.46 & 5.41 \\
GJ 625 & 16:25:24.60 & 54:18:14.80 & 10.40 & 6.06 & 5.83 & 5.60 \\
GJ 659 A & 17:10:10.50 & 54:29:39.80 & 8.80 & 6.23 & 6.12 & 5.97 \\
GJ 659 B & 17:10:12.40 & 54:29:24.50 & 9.29 & 6.13 & 5.97 & 5.83 \\
GJ 684 A & 17:34:59.59 & 61:52:28.39 & 5.23 & 3.89 & 3.74 & \nodata \\
GJ 684 B & 17:34:59.59 & 61:52:28.39 & 8.06 & \nodata & \nodata & \nodata \\
GJ 702 A & 18:05:27.30 & 02:30:00.40 & 4.20 & 2.32 & 2.24 & 2.18 \\
GJ 702 B & 18:05:27.30 & 02:30:00.40 & 6.00 & 3.48 & 3.37 & 3.27 \\
61 Cyg A & 21:06:53.90 & 38:44:57.90 & 5.21 & 2.47 & 2.36 & 2.25 \\
61 Cyg B & 21:06:55.30 & 38:44:31.40 & 6.03 & 3.02 & 2.87 & 2.74 \\
BD+48 3686 & 22:20:07.00 & 49:30:11.80 & 8.57 & 6.58 & 6.51 & 6.45 \\
GJ 860 A & 22:27:59.47 & 57:41:45.15 & 9.59 & 5.04 & 4.78 & \nodata \\
GJ 860 B & 22:27:59.47 & 57:41:45.15 & 10.30 & \nodata & \nodata & \nodata \\
GJ 879 & 22:56:24.10 & -31:33:56.00 & 6.48 & 3.80 & 3.81 & 3.70 \\
HD 220140 A & 23:19:26.60 & 79:00:12.70 & 7.54 & 5.74 & 5.66 & 5.60 \\
HD 220140 B & 23:19:26.60 & 79:00:12.70 & \nodata & \nodata & \nodata & \nodata \\
GJ 896 A & 23:31:52.20 & 19:56:14.10 & 9.95 & 5.24 & 4.99 & 4.64 \\
GJ 896 B & 23:31:52.20 & 19:56:14.10 & 12.40 & 6.98 & 6.68 & 6.28 \\
\enddata
\tablecomments{Coordinates are epoch J2000.0 and are mostly from
\citet{hip}.  $H$ and $K$ magnitudes are from \citet{2MASS},
or else calculated from Simbad website spectral types and 
$V$ magnitudes using Table 7.6 of \citet{aaq}.  $L'$ magnitudes
are similarly calculated from either $V$ or $K$ values.}
\end{deluxetable}
\end{center}

As can be seen from Table \ref{tab:samp01}, some estimates
have placed the ages of some of our stars well below 100 Myr.  We
have chosen to approximate these ages as 100 Myr.
There are several reasons for this.  First, the \citet{bur} 
models we have adopted do not
give the type of observables we need for planets younger
than 100 Myr.  Second, setting the ages of these
stars slightly older than they are thought to be
fits in with our generally conservative approach to
the volatile subject of extrasolar planet searches, and
ensures that our survey results do not hang on just
a few very young stars and will not be invalidated
if the age estimates are revised upward.  Finally,
setting the ages conservatively hedges our results
to some extent against the possibility suggested in
\citet{faintJup} that young massive planets may be
far fainter than expected because much of the gravitational potential
energy of the accreting material may get radiated away
in an accretion shock and thus never get deposited
in the planet's interior.  Figure 4 in \citet{faintJup}
shows that in this accretion scenario planets start
out at much lower luminosities than predicted by
`hot start' models such as those of \citet{bur},
but over time the predictions converge.  By 100 Myr,
the differences are less than an order of magnitude
for planets less massive than 10 \mjup, and are negligible for planets
of 4 \mjup~and lower masses.

\section{Observations and Image Processing} \label{sec:Observe}

\subsection{The Instrument} \label{clio_ins}

The Clio instrument we used for our observations has been well described 
elsewhere (\citet{freed},
\citet{suresh}, and \citet{oldvega}).  We present only a brief overview here.

The MMT AO system delivers a lower thermal background than others because it
uses the world's first deformable secondary mirror, thereby avoiding the multiple warm-mirror
reflections (each adding to the thermal background) that are needed in 
other AO systems.  This unique property makes the 
MMT ideal for AO observations in wavelengths such as the $L'$ and $M$ bands
that are strongly affected by thermal glow.  Clio was developed to take advantage of
this to search for planets in these bands.  It saw first light as a simple
imager offering F/20 and F/35 modes.  The design allowed for coronagrapic
capability, which has since been developed \citep{phaseplate} but was not
used in our survey.  In the F/20 mode, which we
we used for all the observations reported herein, Clio's 
field of view is 15.5$\times$12.4 arcseconds.  Its plate scale 
is $0.04857 \pm 0.00003$ arcseconds per pixel, which gives finer than Nyquist
sampling of the diffraction-limited point spread function 
(PSF) of the MMT in the $L'$ and $M$ bands.

\subsection{Observations} \label{sec:obs}
For each star in our sample we sought to acquire about one
hour or more of cumulative integration at the $L'$ band.  In most cases
we achieved this.  For some of our brightest nearby targets we
acquired $M$ band integrations as well.  If possible
we observed the star through transit, not only to minimize
airmass, but also to obtain the greatest possible amount
of parallactic rotation.  Parallactic rotation is important
because it causes image artifacts from the telescope
to rotate with respect to real sources, rendering them
more distinguishable.  To enhance this effect, we observed
with the instrument rotator off, so that rays and ghosts
from the Clio instrument itself would also rotate, and could
be suppressed by the same procedures that suppressed telescope
artifacts (see Section \ref{sec:datproc}).

After acquiring each target with MMTAO, we determined a
single-frame integration time for our science images
based solely on the sky background.  This integration time was
chosen so that the sky background flux filled $60-80\%$ of
the detector full-well capacity.  This ensured that beyond
the speckle halo of the star the observations
were background-limited rather than readnoise limited.  
The optimal integration time changed due to night-to-night 
variations in sky brightness, usually ranging from
1.5 to 2.0 seconds in $L'$ and from 100-200 msec at $M$;
see Table \ref{tab:obs01} for details.
The science exposures generally saturated the primary star.
When possible, we interleaved a few shorter exposures providing
unsaturated images.  These could be used later to determine
the true PSF delivered by the AO system during observations
of a particular star.
 
In normal operation Clio coadds several individual frames and saves them as a single FITS
image.  We used this option except for our observations of 
the star GJ 380, for which we saved and processed the frames individually.
The increased data volume and processing runtimes for GJ 380
outweighed any minor advantages the single-frame approach 
may offer in terms of image quality.
Coadding delivers good-quality data much more efficiently.

Table \ref{tab:obs01} shows the date on which each of our target stars
was observed, the nominal single-frame integration time, the coadds, 
and the number of coadded FITS images we acquired.  The true single-frame integration
for Clio is the nominal integration plus 59.6 msec.  
Table \ref{tab:obs02} gives the full science integration,
parallactic rotation, and mean airmass for each star.

We took our data using the standard IR imaging technique of nodding,
in which a sequence of images is taken in one position, the
telescopes is moved (`nodded') slightly, and then another image sequence
is acquired.  Images taken at one position can then be subtracted
from images taken at the other position.  In contrast to the
on-source/off-source nodding used in some types of observations,
we place the science target on the detector in both nod positions
to maximize the useful data aquired.  Artifacts of the bright sky 
interacting with the telescope and the detector vanish on nod
subtraction, while real celestial objects, including the target 
star itself, appear as bright and dark images
separated by a distance set by the nod amplitude 
(typically about half our field of view). 
Nodding is a powerful technique, and is practically indispensible for
$L'$ and $M$ band observations.  We typically nodded the 
telescope every 2-5 minutes.  This was short enough that 
alterations in the sky background did not introduce
appreciable noise into our data -- in sharp contrast to, e.g., 10 $\mu$m $N$
band observations, where a `chopping' mirror must be used to 
switch between source and sky on a timescale of seconds or less.  

\begin{deluxetable}{llcrrr}
\tablewidth{0pc}
\tablecolumns{6}
\tablecaption{Observations of Science Targets: Basic Parameters \label{tab:obs01}}
\tablehead{ & \colhead{Date Obs.} & & & & \\ \colhead{Star} & \colhead{(yyyy/mm/dd)} & \colhead{Band} & \colhead{Clio int(msec)} & \colhead{Coadds} & \colhead{\# Images}} 
\startdata
GJ 659A & 2006/04/11 & L' & 2000 & 10 & 90 \\
GJ 354.1A & 2006/04/12 & L' & 2000 & 10 & 232 \\
GJ 450 & 2006/04/12 & L' & 2000 & 10 & 260 \\
GJ 625 & 2006/04/12 & L' & 2000 & 10 & 208 \\
GJ 349 & 2006/04/13 & L' & 2000 & 10 & 240 \\
GJ 564 & 2006/04/13 & L' & 2000 & 10 & 193 \\
GJ 3876 & 2006/04/13 & L' & 2000 & 25 & 68 \\
GJ 3860 & 2006/06/09 & L' & 1500 & 15 & 170 \\
HD 139813 & 2006/06/09 & L' & 1200 & 20 & 148 \\
GJ 702 AB\tablenotemark{a} & 2006/06/09 & L' & 1200 & 20 & 95 \\
61 Cyg A & 2006/06/09 & L' & 1200 & 20 & 133 \\
BD+60 1417 & 2006/06/10 & L' & 1200 & 20 & 160 \\
$\xi$ Boo AB\tablenotemark{a} & 2006/06/10 & L' & 1200 & 20 & 157 \\
61 Cyg B & 2006/06/10 & L' & 1500 & 15 & 140 \\
GJ 519 & 2006/06/10 & L' & 1500 & 15 & 180 \\
BD+48 3686 & 2006/06/11 & L' & 1200 & 20 & 130 \\
$\xi$ Boo AB\tablenotemark{a} & 2006/06/11 & M & 100 & 100 & 260 \\
GJ 684 AB\tablenotemark{a} & 2006/06/11 & L' & 1200 & 20 & 120 \\
GJ 505 AB\tablenotemark{a} & 2006/06/11 & L' & 1200 & 20 & 149 \\
GJ 659 B & 2006/06/12 & L' & 1200 & 20 & 170 \\
61 Cyg A & 2006/06/12 & M & 100 & 100 & 176 \\
GJ 860 AB\tablenotemark{a} & 2006/06/12 & L' & 1200 & 20 & 104 \\
61 Cyg B & 2006/07/12 & M & 100 & 100 & 274 \\
GJ 896 AB\tablenotemark{a} & 2006/07/13 & L' & 1500 & 20 & 105 \\
$\epsilon$ Eri & 2006/09/09 & M & 130 & 100 & 180\\ 
GJ 5 & 2006/09/11 & L' & 1500 & 15 & 210 \\
$\epsilon$ Eri & 2006/09/11 & L' & 1500 & 15 & 184 \\
GJ 117 & 2006/12/01 & L' & 1500 & 15 & 139 \\
GJ 211 & 2006/12/01 & L' & 1500 & 15 & 170 \\
GJ 282 A & 2006/12/01 & L' & 1500 & 15 & 190 \\
HD 1405 & 2006/12/02 & L' & 1500 & 15 & 98 \\
GJ 159 & 2006/12/02 & L' & 1500 & 15 & 180 \\
GJ 216 A & 2006/12/02 & L' & 1500 & 15 & 158 \\
GJ 278 C & 2006/12/02 & L' & 1500 & 15 & 132 \\
GJ 355 & 2006/12/02 & L' & 1500 & 15 & 159 \\
GJ 879 & 2006/12/03 & L' & 1500 & 15 & 54 \\
HD 220140 AB\tablenotemark{a} & 2006/12/03 & L' & 1500 & 15 & 170 \\
GJ 166 BC\tablenotemark{a} & 2006/12/03 & L' & 1500 & 15 & 149 \\
GJ 311 & 2006/12/03 & L' & 1500 & 15 & 90 \\
GJ 410 & 2006/12/03 & L' & 1500 & 15 & 100 \\
$\tau$ Ceti & 2007/01/04 & L' & 1700 & 15 & 160 \\
HD 29391 & 2007/01/04 & L' & 1700 & 15 & 200 \\
BD+20 1790 & 2007/01/04 & L' & 1700 & 15 & 188 \\
HD 96064 AB\tablenotemark{a} & 2007/01/05 & L' & 1700 & 15 & 180 \\
HD 77407 AB\tablenotemark{a} & 2007/01/05 & L' & 1700 & 15 & 79 \\
HD 78141\tablenotemark{b} & 2007/04/11 & L' & 1700 & 15 & 203 \\
HD 113449 & 2007/04/11 & L' & 1500 & 15 & 190 \\
GJ 702 AB\tablenotemark{a} & 2007/04/11 & M & 200 & 100 & 144 \\
GJ 380 & 2007/04/30 & L' & 1500 & 1 & 2066 \\
\enddata
\tablecomments{The `Clio int' column gives the nominal
single-frame integration time for Clio in msec.  The
actual single frame integration is 59.6 msec
longer in every case.}
\tablenotetext{a}{\footnotesize{These stars were sufficiently close binaries that
both stars appeared on the same Clio images, and meaningful sensitivity
to substellar objects could be obtained around both.}}
\tablenotetext{b}{\footnotesize{A small fraction of the images of this star
were accidentally taken with a 1500 msec rather than a 1700 msec nominal
integration time.}}
\end{deluxetable}

\begin{deluxetable}{lccccc}
\tablewidth{0pc}
\tablecolumns{6}
\tablecaption{Observations of Science Targets: Data Acquired \label{tab:obs02}}
\tablehead{\colhead{Star} & \colhead{Band} & \colhead{Exposure(sec)} &
  \colhead{Mean Airmass} & \colhead{Rotation} & \colhead{Proc. Methods}}
\startdata
GJ 659 A & L' & 1853.64 & 1.113 & 15.80$^{\circ}$ & a, b, d, e \\ 
GJ 354.1 A & L' & 4778.27 & 1.032 & 130.75$^{\circ}$ & a, b, d, e, x, y \\ 
GJ 450 & L' & 5354.96 & 1.031 & 110.37$^{\circ}$ &  a, b, d, e, x, y \\ 
GJ 625 & L' & 4283.97 & 1.117 & 45.65$^{\circ}$ & a, b, d, e, x, y \\ 
GJ 349 & L' & 4943.04 & 1.178 & 40.61$^{\circ}$ & a, b, d, e \\ 
GJ 564 & L' & 3975.03 & 1.036 & 70.70$^{\circ}$ & a, b, d, e \\ 
GJ 3876 & L' & 3501.32 & 1.601 & 27.23$^{\circ}$ &  a, b, d, e \\ 
GJ3860 & L' & 3976.98 & 1.086 & 47.09$^{\circ}$ & a, b, d, e \\ 
HD139813 & L' & 3728.42 & 1.529 & 30.15$^{\circ}$ &  a, b, d, e \\ 
GJ 702 AB\tablenotemark{a} & L' & 2393.24 & 1.149 & 25.50$^{\circ}$ & a, b,
d, e, f, g \\ 
61 Cyg A & L' & 3350.54 & 1.012 & 101.25$^{\circ}$ & a, b, d, e \\ 
BD+60 1417 & L' & 4030.72 & 1.153 & 37.65$^{\circ}$ & a, b, d, e \\ 
$\xi$ Boo AB\tablenotemark{a} & L' & 3955.14 & 1.047 & 71.20$^{\circ}$ & a, b,
d, e, f, g \\ 
61 Cyg B & L' & 3275.16 & 1.012 & 103.68$^{\circ}$ & a, b, d, e \\ 
GJ 519 & L' & 4210.92 & 1.011 & 139.97$^{\circ}$ & a, b, d, e \\ 
BD+48 3686 & L' & 3274.96 & 1.074 & 35.97$^{\circ}$ & a, b, d, e \\ 
$\xi$ Boo AB\tablenotemark{a} & M & 4149.60 & 1.060 & 46.142$^{\circ}$ & a, b,
d, e, f, g \\ 
GJ 684 AB\tablenotemark{a} & L' & 3023.04 & 1.175 & 24.15$^{\circ}$ & d, e, y,
g \\
GJ 505 AB\tablenotemark{a} & L' & 3753.61 & 1.070 & 45.30$^{\circ}$ & a, b, d,
e, f, g, x, y \\ 
GJ 659 B & L' & 4282.64 & 1.112 & 43.93$^{\circ}$ & a, b, d, e \\ 
61 Cyg A & M & 2808.96 & 1.025 & 44.24$^{\circ}$ & a, b, d, e \\ 
GJ 860 AB\tablenotemark{a} & L' & 2619.97 & 1.133 & 24.55$^{\circ}$ & a, d, e,
g, y \\
61 Cyg B & M & 4373.04 & 1.018 & 118.96$^{\circ}$ & d, e, y \\
GJ 896 AB\tablenotemark{a} & L' & 3275.16 & 1.026 & 66.49$^{\circ}$ & a, b, d,
e, f, y \\ 
$\epsilon$ Eri & M & 3412.80 & 1.334 & 23.406$^{\circ}$ & d, e, y \\ 
GJ 5 & L' & 4912.74 & 1.011 & 146.98$^{\circ}$ & a, b, d, e, x, y \\ 
$\epsilon$ Eri & L' & 4304.50 & 1.342 & 36.92$^{\circ}$ & d, e, y \\ 
GJ 117 & L' & 3251.77 & 1.463 & 34.05$^{\circ}$ & a, b, d, e, x, y \\ 
GJ 211 & L' & 3976.98 & 1.097 & 50.12$^{\circ}$ & a, b, d, e, x, y \\ 
GJ 282 A & L' & 4444.86 & 1.281 & 30.28$^{\circ}$ & a, b, d, e, x, y \\ 
HD 1405\tablenotemark{b} & L' & 2292.61 & 1.036 & 162.97$^{\circ}$ & a, b, d,
e, x, y \\ 
GJ 159 & L' & 4210.92 & 1.189 & 37.65$^{\circ}$ & a, b, d, e, x, y \\ 
GJ 216 A & L' & 3696.25 & 1.739 & 30.10$^{\circ}$ & a, b, d, e, x, y \\ 
GJ 278 C\tablenotemark{b} & L' & 3088.01 & 1.017 & 170.627$^{\circ}$ & a, b, c, d, e, x, y \\ 
GJ 355 & L' & 3719.65 & 1.380 & 25.74$^{\circ}$ & a, b, c, d, e, x, y \\ 
GJ 879 & L' & 1263.28 & 2.232 & 11.68$^{\circ}$ & a, c, d, x, y \\ 
HD 220140 AB\tablenotemark{a} & L' & 3976.98 & 1.494 & 14.14$^{\circ}$ & a, b,
d, e, f, g, x, y \\ 
GJ 166 BC\tablenotemark{a} & L' & 3485.71 & 1.301 & 28.66$^{\circ}$ & a, b, d,
e, x, y \\ 
GJ 311 & L' & 2105.46 & 1.201 & 26.23$^{\circ}$ & a, b, c, d, e, x, y \\ 
GJ 410 & L' & 2339.40 & 1.026 & 34.26$^{\circ}$ & a, b, c, d, e, x, y \\ 
$\tau$ Ceti & L' & 4223.04 & 1.535 & 37.03$^{\circ}$ & a, b, d, e, x, y \\ 
HD 29391 & L' & 5278.80 & 1.227 & 39.63$^{\circ}$ & a, b, c, d, e, x, y \\ 
BD+20 1790 & L' & 4962.07 & 1.068 & 47.94$^{\circ}$ & a, b, d, e, x, y \\ 
HD 96064 AB\tablenotemark{a} & L' & 4750.92 & 1.252 & 41.74$^{\circ}$ & a, b,
d, e, x, y \\ 
HD 77407 AB\tablenotemark{a} & L' & 2085.13 & 1.008 & 95.44$^{\circ}$ & a, b,
c, d, e, f, g, x, y \\ 
HD 78141\tablenotemark{c} & L' & 5297.98 & 1.022 & 109.11$^{\circ}$ & a, b, c,
d, e, x, y \\ 
HD 113449 & L' & 4444.86 & 1.263 & 35.36$^{\circ}$ & a, b, d, e, x, y \\ 
GJ 702 AB & M & 3738.24 & 1.171 & 32.70$^{\circ}$ & d, e, g, y \\
GJ 380 & L' & 3222.13 & 1.341 & 20.58$^{\circ}$ & a, b, d, e, x, y \\ 
\enddata
\tablecomments{Proc. Methods refers not to the data that were
acquired, but to different methods used in processing the
data.  Each method represents
a different master image produced by stacking the entire
data set after applying a particular set of pre-stack
processing algorithms.  For example, four separate
master images were made of the star GJ 659A.  Each
was a stack of all the images acquired, processed
using a different method of pre-stack image processing: the
`a', `b', `d', and `e' methods in the case of this star.
The different processing methods are explained in Section \ref{sec:datproc}}
\tablenotetext{a}{\footnotesize{These stars were sufficiently close binaries that
both stars appeared on the same Clio images, and meaningful sensitivity
to substellar objects could be obtained around both.}}
\tablenotetext{b}{\footnotesize{Though the rotation on this star is very large, difficulties
arise because the star transited very near the zenith and almost all the rotation
happened in a short span of time during which observations were not possible.
PSF subtraction had to be performed on a subset of the data with equal numbers
of images on each side of transit.}}
\tablenotetext{c}{\footnotesize{A small fraction of the images of this star
were accidentally taken with a 1500 msec rather than a 1700 msec nominal
integration time.  The total exposure time has been corrected accordingly.}}
\end{deluxetable}

\subsection{Image Processing} \label{sec:datproc}
Image processing for AO planet search images tends
to be complex and sophisticated.  We have given a
brief outline of our processing pipeline in \citet{newvega},
which is applicable to the current work, and we hope
to detail the unique aspects of our pipeline in a separate
future paper.  Here we will briefly describe the processing
sequence, stressing aspects that were not covered in
\citet{newvega}, but which become more important for the
larger set of stars, processed over a longer period of time,
that we describe herein.

\begin{figure*}
\plotone{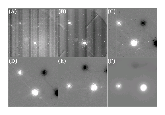}
\caption[Image Processing Sequence Applied to GJ 896AB]{\textbf{(A)}
Raw image of the nearby binary star GJ 896.  \textbf{(B)} Same
image after dark subtraction and flatfielding.  Contrast
stretched 5$\times$ relative to (A).  \textbf{(C)} Same image
after nod subtraction.  Contrast stretched 2.5$\times$ relative
to (B).  \textbf{(D)} Same image after correction for bad pixels
and bad columns.  \textbf{(E)} Same image after shifting and rotation.
\textbf{(F)} Final stack made from 105 images like (E).  Unsharp
masking has not yet been applied.  The field of view for each
tile is 10.6 arcsec square.}
\label{fig:procsteps}
\end{figure*}

We begin the processing of each Clio image by normalizing 
it to a single coadd, subtracting an equal-exposure dark
frame usually taken immediately before or after the science data
sequence, and dividing by a flat frame.  There follows an initial step
of bad-pixel fixing.  The next step is nod subtraction:
from every image we subtract an identically processed
copy of an image from the opposite nod position.  This nod subtraction image
is scaled (by a factor that is always very close to unity)
so that its mean sky brightness exactly matches that of the
science image from which it is being subtracted; 
the scaling is useful to compensate for small
variations in sky brightness.  Further bad-pixel fixing
and bad-column correction follows.  Finally, an algorithm to
remove residual pattern noise is applied, and the image
is zero-padded, shifted, and rotated in a single bicubic
spline operation so that celestial north is up and the
center-of-mass centroid of the primary star is located in the exact
center of the image.  See Figure \ref{fig:procsteps}
for an example of our processing sequence, applied to
the nearby binary star GJ 896.

The rotation places celestial north up on the images with an
accuracy of about 0.2 degrees.  Since we do not use the instrument
rotator, a different rotation is required for each image: the 
parallactic angle plus a constant offset, which we
determine by observing known binary stars (this is further
described in section \ref{sec:bin}).  While parallactic 
rotation of bright binary stars over just tens of seconds has been 
detected due to the high internal precision of Clio astrometry, 
in no case does sufficient parallactic rotation occur during a 
Clio coadd sequence to appreciably blur the science images.

We have confirmed that the clean, symmetrical stellar images produced by the
MMT AO system at the $L'$ and $M$ bands give accurate, consistent 
center-of-mass centroids even if saturated. This is important
for our survey since our pre-stack registration of images is
based in most cases on centroids of a saturated primary.  If
the variation in such centroids is more than about one pixel,
faint sources will be substantially blurred in the final stacks,
and our point-source sensitivity will be appreciably reduced.
In practice, however, we find that faint sources (and bright
secondaries in binary systems) do in fact appear sharp in our
image stacks. Images we took of Procyon (unpublished) and
of 61 Cyg A and B (see Figures \ref{fig:comp4} and \ref{fig:comp5})
illustrate this in an especially striking manner, because our images
of Procyon were more severly saturated than any reported herein,
while our 61 Cyg A and B images were among the most saturated in
our survey. In all three of these cases, sharp images of faint 
companions (the orbiting white dwarf in the case of Procyon; background
stars in the cases of 61 Cyg A and B) appeared in the final image 
stacks, which were registered solely based on center-of-mass 
centroids of the heavily saturated primary.  The consistency
of such centroids is confirmed to an even tighter tolerence based
on our observations of binary survey targets, in which both
saturated and unsaturated images were aquired.  For example, the total
differences between our saturated and unsaturated astrometry at $M$ band
for the binary stars GJ 702 and $\xi$ Boo were only 0.0007 arcsec 
and 0.0039 arcsec, respectively (where differences in separation 
and position angle have been combined).
The same saturated vs. unsaturated differences for our $L'$ astrometry
of the binary stars $\xi$ Boo, HD 77407, GJ 505, and GJ 166BC were
0.0088 arcsec, 0.0038 arcsec, 0.0026 arcsec, and 0.0015 arcsec, respectively.
These values are based on averages of astrometric measurements performed
on individual frames prior to stacking.  The internal scatter in
the astrometry of saturated images was also very low, even though
the saturated measurements spanned about an hour of time and tens of
degrees of parallactic rotation in each case, giving ample opportunity
for any defects in the saturated astrometry to manifest themselves.  
In all cases tested, center-of-mass centroids of saturated images are 
self-consistent, and consistent with centroids of unsaturated images,
to considerably greater precision than necessary for the purposes
of our survey.

We stack our processed images to make a master image for each
processing method using a creeping mean combine.
This method of image stacking uses a single parameter, the
rejection fraction, which we set to 20\% for our standard master
images.  The mean of each given pixel through
the image stack is computed, the most deviant value is rejected,
and the mean is computed again.  This procedure is iterated
until the required fraction of data points have been rejected.  
One of us (S. S.) developed an $N\log(N)$ implementation
that greatly improved the speed of our processing pipeline.  We chose
the creeping mean over the more commonly used median
with sigma-clipping because the creeping mean can deliver cleaner
final stacks when, as with Clio, the raw images contain bright,
slowly-rotating ghosts and diffraction rays.  In clean
sky away from all ghosts and rays, the median delivers slightly
lower rms noise, since it rejects fewer data points.

Our final stacked images contain dark, high-noise regions
on either side of each bright star, due to the negative
star images from nod subtraction.  Since we usually keep
a constant nod direction referenced to the telescope,
for data sets with significant parallactic
rotation the dark regions are spread into arcs and
weakened by the creeping mean stack.  To 
further alleviate the dark regions and to enhance the visibility
of faint point sources against the bright stellar halo
itself, we unsharp mask the final, stacked images.  We do
this by convolving the image
with a Gaussian kernel of $\sigma = 5$ pixels, and then subtracting
this convolved version from the original image.  The 
full width at half maximum (FWHM) of the
Gaussian kernel is 11.8 pixels, as opposed to a FWHM of about 3 pixels 
for a typical PSF, so the unsharp masking does not strongly
reduce the brightness of real point sources.  This step
marks the end of our image processing pipeline.

The above describes our baseline processing method.
We developed six specializations of this method, which we
call the `b,' `c,' `d,' `e,' `x,' and `y' processing strategies,
while the previously described baseline method itself is called `a'.  
The data from each star in our survey were processed several times,
each time using a different one of these specialized methods,
and each producing a separate master image.  Having multiple
master images based on different processing methods is
helpful because the different methods enhance sensitivity
to planets in different parts of the images, and because the
master images from different methods provide a quasi-independent
check on the reality of suspected faint sources.  We will now
describe how these different specialized processing methods
function.

In the `b'
processing method, we suppress the stellar PSF to increase
our sensitivity to faint companions.
To do this, we take advantage of the fact that long-lived
PSF artifacts in stellar images from AO-equipped telescopes tend to
remain fixed with respect to the telescope and/or instrument \citep{sspeckle}.  
When observing with the instrument-rotator off, as we do, real sources
slowly rotate with respect to artifacts as the telescope tracks.  
Science images must be digitally rotated before stacking, as 
described above.  However,
if a stack of \textit{un}-rotated frames is made, a clear image
of the instrumental PSF is obtained, while any real
sources are strongly attenuated by the creeping mean.  We
subtract a properly registered version of such a PSF image
from every science frame prior to final rotation and stacking,
a technique called ADI \citep{marois}.  In our specific
implementation of ADI, we split the image set into a first
and second half, and a PSF image is made using a 50\% rejection
creeping-mean stack of each half.  The PSF image from the second half of
the data is subtracted from every image in the first half,
while the PSF image from the first half of the data is subtracted
from the images in the second half.  The result is powerful 
attenuation of the stellar PSF and greatly increased sensitivity
to close-in companions.  Since parallactic angle changes
monotonically with time in all our observing sequences,
splitting the data into first and second halves helps prevent 
real companions from being partially subtracted
due to appearing at a residual level in the PSF images.  For
stars with insufficient parallactic rotation, very close-in
companions can still be partially subtracted, but a characteristic
dark-bright-dark signature is created which is very
noticeable for companions of sufficient brightness.  However, in
our sensitivity analyses, we have conservatively set the
sensitivity of ADI images to zero inside the radius where
such ADI self-subtraction first becomes significant.

In the `c' reduction method, an azimuthally smoothed
version of the primary PSF is subtracted from the image.
The smoothing is done using creeping-mean rejection in 
a sliding annular arc centered on the primary, with
parameters set so that real sources vanish essentially
completely from the smoothed PSF and therefore cannot
be dimmed in the subtraction.  The quality of PSF subtraction
achieved is usually substantially inferior to the `b' method,
and the `c' method is therefore used relatively seldom.
Sometimes it is employed because insufficient parallactic
rotation renders the `b' method less useful, or because the `c' image
with its different speckle pattern is desired as a 
quasi-independent check on candidate sources detected in
the `b' method image.

In the `d' reduction method, each image is unsharp
masked \textit{before} the stack.  The final stacked
image is unsharp masked again.  While unsharp masking
is a linear process, the creeping mean stack is not, so the
results are different from simply unsharp-masking twice
after the final stack.  This is especially significant
for bright stars with intense seeing halos.  Due to our
nod subtraction method, it often happens for such stars that
a given x,y pixel location falls on a bright, positive
seeing halo for images taken in the first nod position,
and on a negative, subtracted seeing halo for images taken in
the other nod position: that is, the statistics through
the image stack at this pixel location are strongly bimodal.  Under such
circumstances the creeping mean will settle on either the
positive or the negative side of the distribution -- and which
one it settles on can be different for adjacent pixels.
This causes intense `bimodality noise' that is essentially
an artifact of the stack.  Note that using a median stack
instead will not necessarily fix the problem, as there
may well be no middle ground between the positive and
negative seeing halos.  Pre-stack unsharp masking removes
the seeing halos and thus resolves the problem of bimodality
noise, enormously improving the quality of the final stacked
image for bright stars.  For fainter stars, the results
are more similar to simply unsharp-masking the final image
twice.  However, the specific noise pattern is substantially
changed, which can aid in confirming faint sources: the `d' method
image can provide a quasi-independent confirmation for a faint
source marginally detected in the baseline `a' image.  The `e' data 
reduction method combines the `b' and `d' methods: ADI
is applied, and then the pre-stack unsharp masking is performed.

\begin{figure*}
\plotone{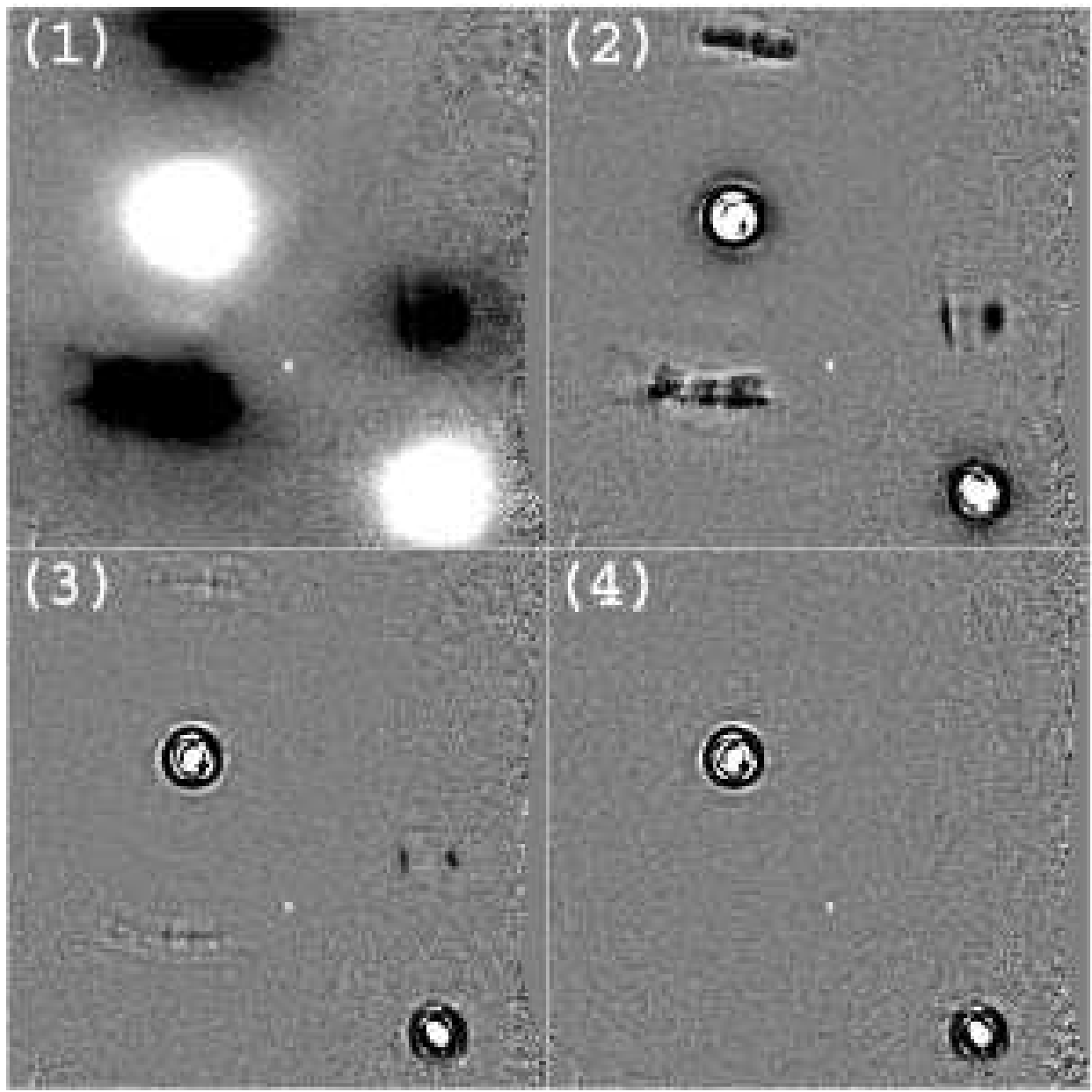}
\caption[Different Processing Methods Applied to HD 96064]{
Different processing methods applied to the wide binary star
HD 96064.  The brightness of the faint source is 
$L' = 13.7$, corresponding to a mass of about 20 \mjup~if
it were a true companion -- however, it is confirmed
to be a background star. \textbf{(1)}
Result of baseline processing (`a' method) before final
unsharp mask.  \textbf{(2)} The `a' method image after unsharp
masking.  Dark nod-subtraction artifacts are somewhat reduced but remain
prominent.  \textbf{(3)} Same data set processed with the `d'
method.  Nod artifacts are greatly
reduced, but still exist as high-noise regions where faint
sources could not be detected.  \textbf{(4)} Same data set
processed with the `y' method.  The nod artifacts are eliminated.
Field shown in each tile is 17 arcsec square.}
\label{fig:ymethod}
\end{figure*}

The `x' data reduction method uses a variant on nod
subtraction that avoids the dark negative images.  Two
master sky images are made, by combining the 
star-free portions of all images in the first and second
halves of the data set.  One of these
star-free master sky images is then subtracted from each 
individual science image in lieu of the ordinary nod subtraction.
To avoid subtracting real sources, the sky image from
the second half of the data set is subtracted from images
in the first half, and vice-versa.
The usefulness of this processing method varies
enormously from one data set to another.  If the sky
background was very stable, the `x' method final image
is almost indistinguishable from that of the baseline `a' method, so that
blinking the two gives the impression that the dark
nod-subtraction artifacts magically disappear.  
If the sky background was highly variable, 
the `x' images are useless due to intense pattern noise.  
The `y' image reduction method is a combination of the `x'
and `d' methods, in which the images are unsharp
masked after the subtraction of the master sky image
but before the final stack.  Figure \ref{fig:ymethod} compares
the results of the `a' method (before and after
the final unsharp masking step), the `d' method, and
the `y' method.  The star is HD 96064, a binary system
in which the secondary is itself a close binary.  A faint
additional companion is also detected, but is confirmed
based on proper motion and $K_S - L'$ color to be a background star
rather than a substellar companion.

\begin{figure*}
\plotone{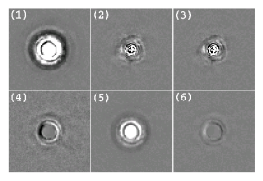}
\caption[Different Processing Methods Applied to GJ 896AB]{\textbf{(1)}
Baseline `a' method final image of GJ 896A.  \textbf{(2)} Same
data set processed with ADI (`b' method).  \textbf{(3)} Same data
set processed with ADI and pre-stack unsharp masking (`e' method).  
\textbf{(4)} Same data set processed with binary star subtraction.  
Background noise is increased because the secondary had to be scaled
up to match the brightness of the primary.  \textbf{(5)} Same data
set, but now showing the `a' method image of the secondary,
rather than the primary.  \textbf{(6)} Same data set, again
showing the secondary, but now processed with binary star
subtraction.  The background is very clean since the primary
was scaled down to subtract away the secondary.  Field shown in
each tile is 3.9 arcsec square.}
\label{fig:bmethod}
\end{figure*}

Two additional processing methods could be applied to
binary stars of near-equal brightness for which both
components appeared on each Clio frame.  A scaled 
version of the PSF of each star could be used
to subtract the other, on a frame-by-frame basis, 
prior to the final stack.  The resulting PSF
subtraction was substantially better than ADI.
We labeled this reduction method `f.'  A version
that also included pre-stack unsharp masking was called `g'.
Figure \ref{fig:bmethod} illustrates our different PSF
subtraction methods, both ADI and binary star subtraction,
as applied to the binary star GJ 896, which was also shown
in Figure \ref{fig:procsteps}.

We applied the `a,' `b,' `d,' and `e' processing methods
to almost all of our stellar data sets, except a very few
for which there was insufficient parallactic rotation to
use the ADI methods without subtracting real sources.
In many instances we also applied the `x' and `y' methods.
We applied the `f' and `g' methods to every binary star
where they would work.  

The methods involving pre-stack
unsharp masking (`d,' `e,' `y,' and `g') always gave
cleaner images, but we used the other methods as well because
pre-stack masking slightly dimmed point sources 
(by about 3-10\%, depending on the AO-corrected
FWHM), and there was a slight chance
this could cause a discovery to be missed.  Our pattern-noise correction
method also dimmed faint point sources by about 15-18\%, based
on tests.  Near the end of our processing, one of us
(M. K.) developed a superior pattern-noise correction that caused
zero dimming, and we also developed a type of unsharp masking
that produced zero dimming to within the measurement error
of our tests.  Only the stars $\epsilon$ Eri ($L'$ and $M$ band),
GJ 684 A, GJ 684 B, GJ 702 A ($M$ band only), GJ 702 B ($M$ band only),
61 Cyg B ($M$ band only), GJ 860 A, and GJ 860 B were
processed using these improvements.  For these stars, 
only the `d,' `e,' `y,' and, where applicable, the 
`g' processing methods were used, since the downside
of pre-stack unsharp masking had been eliminated.

\section{Sensitivity Analysis} \label{sec:sensanal}

\subsection{Sensitivity Estimators} \label{sec:estim}

Our survey arrived at a null result: no planets were
detected.  Our science results, like those of previous
surveys \citep{masciadri,kasper,biller1,GDPS,chauvin}, therefore
take the form of upper limits on the abundance
of extrasolar planets.  The accuracy of such an upper limit
depends entirely on having a good metric for the
sensitivity of the survey observations.

A sensitivity estimator must translate some measurable statistic
of an image into a realistic point-source detection limit.
A procedure which has often been used (see for example
\citet{biller1} and \citet{chauvin}) involves calculating the single-pixel
RMS standard deviation ($\sigma_{pix}$) in different regions on an image, and
adopting a factor (often taken to be 5.0) by which the peak
of a point-source image must exceed this $\sigma_{pix}$ to be
cleanly detected.  All that remains is to map the $5 \sigma_{pix}$
PSF peak to a magnitude (or $\Delta$-mag), and assign this as
the sensitivity in the image region under consideration.  \citet{biller1}
and \citet{kasper}, among others, have discussed possible choices
for the size and shape of the regions over which $\sigma_{pix}$
is calculated, with the objective of obtaining smooth and accurate
plots of point-source sensitivity vs. separation from the star.

While the method above produces excellent results when correctly
applied, we sought to adopt a slightly more sophisticated approach.
One reason for this is that calculating sensitivity based on comparing
the single-pixel RMS to the peak of the PSF does not take into account
the FWHM of the PSF.  If the PSF is several pixels
wide, detection need not depend on the peak height alone: pixels other
than the central peak contain additional flux that can in principle be used to
detect the point source at a lower peak flux than would be possible
for a narrower PSF.  We have explored three possible sensitivity
estimation methods that attempt to consider all the flux contained
in the image of a point source, rather than only the peak of the PSF.
The first solution we considered was calculating $\sigma_{pix}$ just
as for the previous method, and then translating this to a detection limit using 
simple $\sqrt{n}$ statistics:

\begin{equation} \label{sens1}
\sigma_{PSF} = \sigma_{pix} \sqrt{\pi r^2} = \sigma_{pix} r \sqrt{\pi}.
\end{equation}

Where $\sigma_{PSF}$ is the PSF-scale noise in the image, $\sigma_{pix}$
is the single-pixel RMS as before, and $r$ is the radius of the image of a
point source (i.e. about half the FWHM of the PSF).  Since
not all the flux of a real point source will fall within the aperture
of radius $r$, an aperture correction must be applied as a final
step.  Then, for example, the 5$\sigma$ point-source sensitivity
will be 5 $\sigma_{PSF}$ times the aperture correction.  This sensitivity
limit would represent an actual integrated flux, which could be converted
directly to magnitudes using a photometric calibration. 
We will call this sensitivity estimation technique `Method 1'.

The simple $\sqrt{n}$ statistics
used in Method 1 assume that the brightness of each pixel
is a random variable independent of its neighboring pixels:
that is, that the noise is spatially uncorrelated.
This assumption is violated for speckle residuals close to a star,
and for a host of other stellar artifacts that are present in AO images
(ghosts, diffraction rays, etc.).  We have confirmed by careful
tests that in the presence of speckle noise, Method 1 overestimates
the true point-source sensitivity by up to 0.9 magnitudes.  This
applies to a good implementation of the method in which $\sigma_{pix}$
is calculated over image regions spanning many PSF sizes.  When
the statistics region used is too small, the sensitivity will be
overestimated even more.

The problem with Method 1 is that clumps of correlated bright
or dark pixels introduce more PSF-scale noise into the image
than can be predicted from the single-pixel RMS.  \citet{GDPS}
solved this problem by convolving their image with a circular
disk of radius $r$, effectively summing up the brightness 
within many small circular apertures at this radius, one
aperture centered on each pixel throughout the image.  Then $\sigma_{PSF}$
will simply equal the RMS variation of the aperture sums
(that is, of the convolved image).  This is sensitivity estimation
by aperture photometry of the noise background.  
As with previously discussed methods, it is important to 
calculate the statistic over an image region large
enough to contain many PSFs.  In our implementation, the region
over which the statistic $\sigma_{PSF}$ is calculated is either 
a disk of 8 pixel radius, or, close to the star, an annular arc 
one pixel wide and 45 pixels long, at constant radius from the 
star. For simplicity, we will sometimes refer to this 
\citet{GDPS} method as `Method 2'.  As with Method 1, an aperture
correction must be applied as a final step.

Method 3 has already been described in \citet{newvega}.  
It is analagous to Method 2, but rather
than performing aperture photometry centered
on every pixel of the image, one performs PSF-fitting photometry.
If the PSF has been properly normalized,
no aperture correction is necessary for this method.
We used PSF images from the short, unsaturated
exposures described in Section \ref{sec:obs}.

In tests using our own real data, we find that
the \citet{GDPS} method and Method 3 agree to
within reasonable uncertainty everywhere,
while Method 1 agrees with the other two only in
regions of very clean sky.  Method 1 overestimates
the sensitivity by about 0.2 magnitudes in the presence
even of very faint ghost residuals, and by about 0.9
magnitudes in the strong residual speckle noise close
to the star.  Herein, as in \citet{newvega}, we have used Method 3
for our final sensitivity maps.  It seemed slightly more
conservative close to the star than Method 2, though, again, our
tests showed no significant difference between Method 3 and the
method of \citet{GDPS}.
Far from the primary star, the region we use for calculating
the sensitivity statistic is a disk of radius 8 pixels 
(0.39 arcsec, or about 3 $\lambda/D$): that is, large enough to span many 
PSF-sizes, but small enough to sample the local noise properties.  
Close to the star (that is, within 60 pixels or 2.9 arcsec),
we use instead an arc 45 pixels (2.2 arcsec) long
and 1 pixel wide, at a fixed radius from the star.  These
disks or arcs are centered in turn on every pixel of each
image, with the calculated statistics forming a sensitivity
map.

\subsection{Sensitivity Obtained} \label{sec:realsens}

After making a sensitivity map from the stacked image produced
by each processing method applied to the data from a given
star, we apply a slight smoothing to the different
maps, and then combine them into a single master sensitivity
map.  They are combined such that the master sensitivity image 
shows at each location the best sensitivity obtained at that 
location by any processing method that was applied.  
We quote 10$\sigma$ sensitivities: that is, the point source 
sensitivity is ten times the $\sigma_{PSF}$ statistic from Method 3.
10$\sigma$ is chosen as a nominal detection threshold because
we have over 95\% completeness for 10$\sigma$ sources, with
considerably less for 5 or 7$\sigma$ (see Section \ref{sec:blind}).

Our background-limited 10$\sigma$ sensitivity for one-hour
exposures under fair conditions is typically $L'=16.0$, or
$M=13.0$.  Since we can detect some sources down to 5$\sigma$
significance, this corresponds to some chance of finding
objects as faint as $L'=16.75$ or $M=13.75$.  For exposures
longer than an hour or under very good conditions, our background
limited 10$\sigma$ sensitivity ranged as high as $L'=16.5$ or $M=13.3$.
Our median 10$\sigma$ sensitivities close to the stars
were about $\Delta$mag$=6.0$ at 0.5 arcsec and $\Delta$mag$=8.7$
at 1.0 arcsec, though the values could range as high as 7.2 and
9.8, respectively.  The $\Delta$mag values obtained
by shorter wavelength AO observations (e.g. \citet{biller1}
and \citet{GDPS}) are much better due to the smaller diffraction
disk at these wavelengths, but this effect is substantially
compensated by the more favorable planet/star flux ratios
at the $L'$ and $M$ bands.  See \citet{newvega} for a detailed
comparison of the efficacies of different wavelengths for
planet detection in the specific cases of Vega and $\epsilon$ Eri.

\begin{figure*}
\plotone{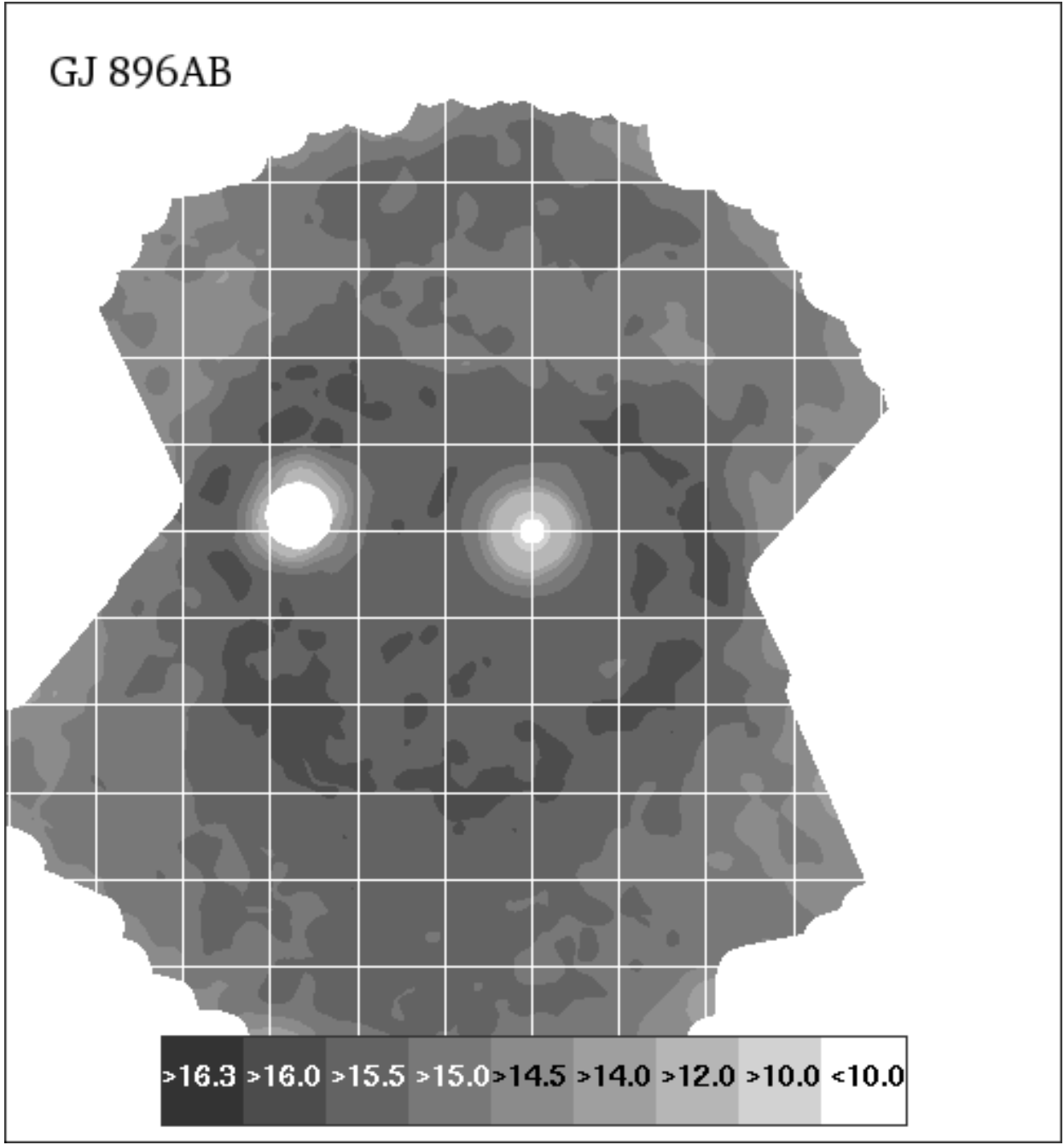}
\caption[Sensitivity Contour Map for GJ 896 AB]{Final
sensitivity contour map for the binary star GJ 896 AB.
10$\sigma$ sensitivities from our Method 3 estimator
are presented, converted to apparent $L'$ magnitudes.
The grid squares superposed for astrometric reference
are 2$\times$2 arcsec.  The darkest contour from
the colorbar is not
present as the 10$\sigma$ sensitivity in this data
set never exceeded $L'=16.3$.} 
\label{fig:GJ896cons}
\end{figure*}

\begin{figure*}
\plotone{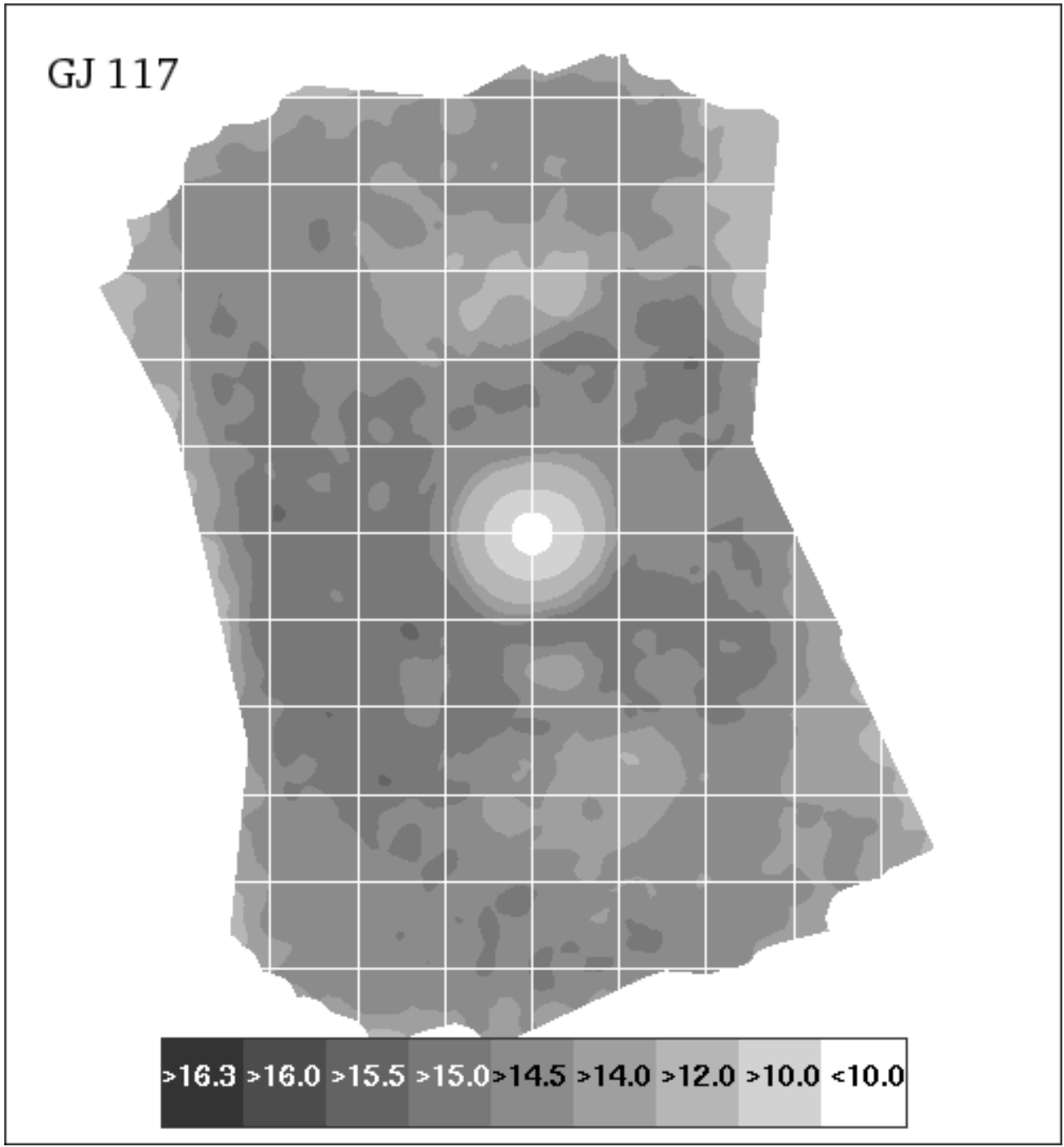}
\caption[Sensitivity Contour Map for GJ 117]{Final
sensitivity contour map for the star GJ 117.
10$\sigma$ sensitivities from our Method 3 estimator
are presented, converted to apparent $L'$ magnitudes.
The grid squares superposed for astrometric reference
are 2$\times$2 arcsec, with the primary star in the
figure's center.  The darkest two contours from
the colorbar are not present as the 10$\sigma$ sensitivity 
in this data set never exceeded $L'=16.0$.} 
\label{fig:GJ117cons}
\end{figure*}

\begin{figure*}
\plotone{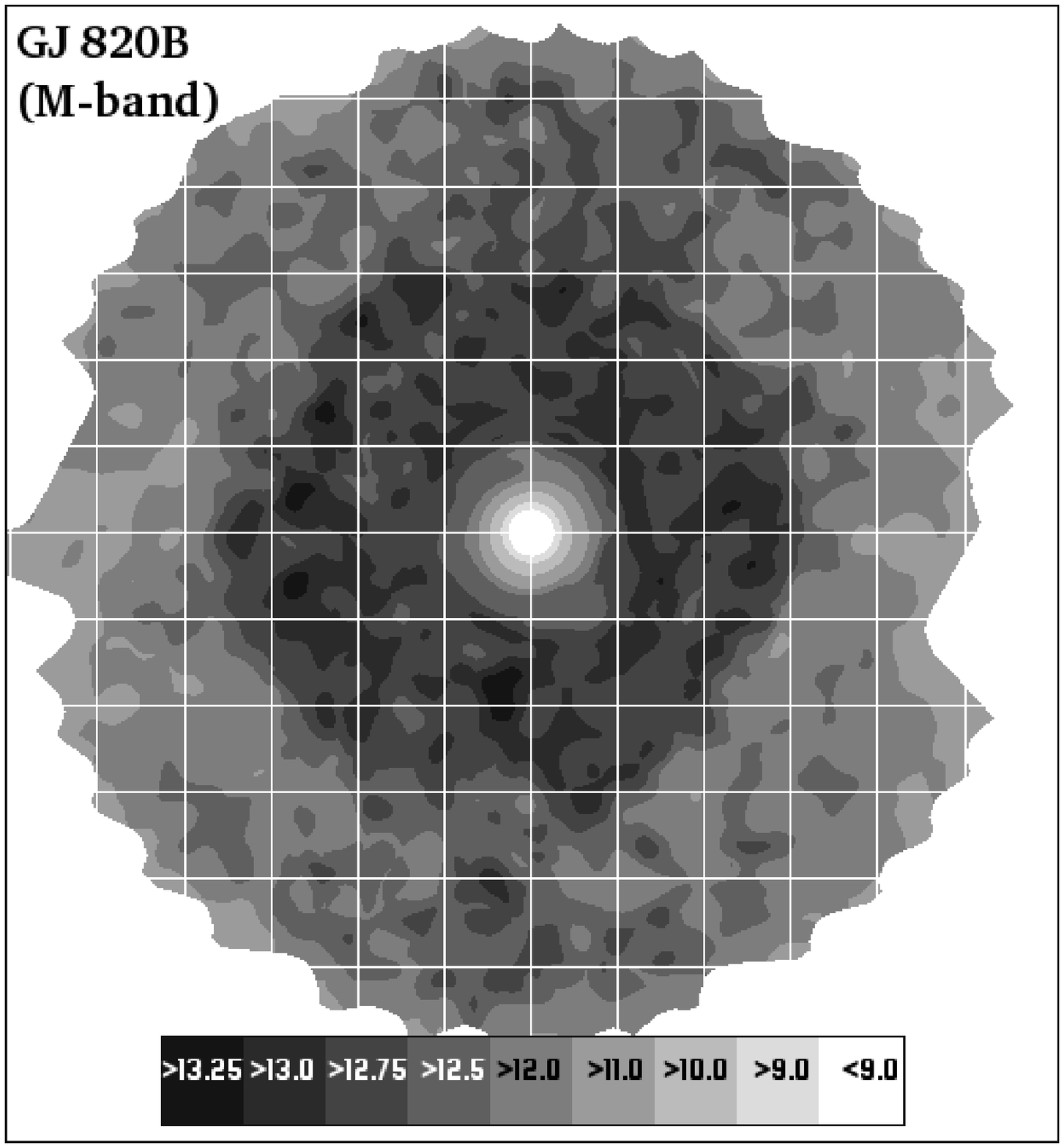}
\caption[Sensitivity Contour Map for 61 Cyg B ($M$ band)]{Final
sensitivity contour map for our $M$ band
observations of the star 61 Cyg B (GJ 820 B).  10$\sigma$ 
sensitivities from our Method 3 estimator
are presented, converted to apparent $M$ band magnitudes.
The grid squares superposed for astrometric reference
are 2$\times$2 arcsec.} 
\label{fig:61CygBcons}
\end{figure*}

Figures \ref{fig:GJ896cons}, \ref{fig:GJ117cons}, and \ref{fig:61CygBcons}
give example sensitivity contour maps for our $L'$ observations
of GJ 896 and GJ 117, and our $M$ band observations of
61 Cyg B, respectively, with 10$\sigma$ sensitivities given
in apparent magnitudes.  Figures of this type for all the
stars observed in our survey can be downloaded from
\url{http://www.hopewriter.com/Astronomyfiles/Data/SurveyPaper/}.

\begin{figure*}
\plottwo{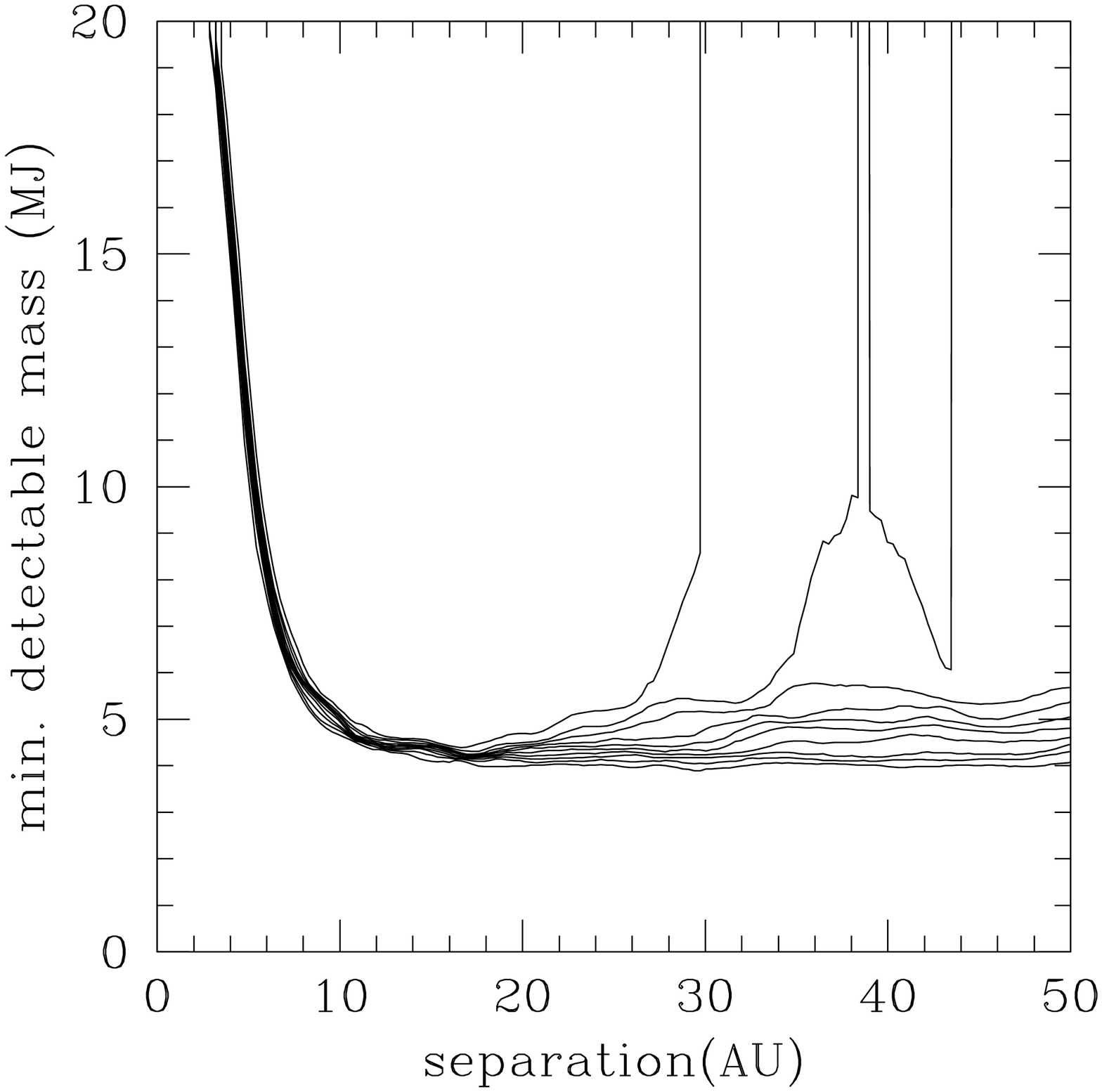}{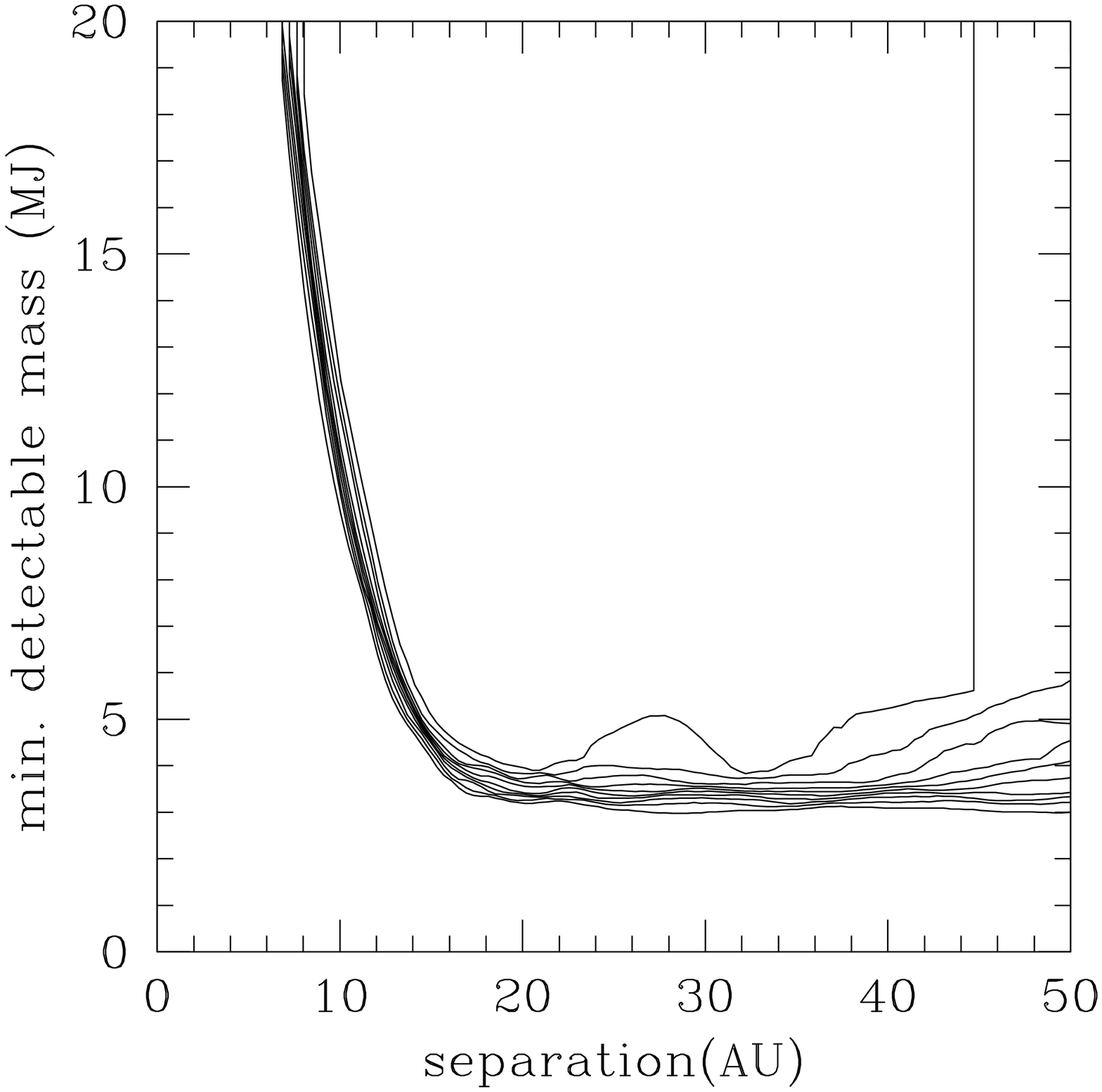}
\caption[Sensitivity Plots for GJ 896 A and GJ 117]
{Minimum detectable
planet mass vs. projected separation in AU for GJ 896 A (left)
and GJ 117 (right).  10$\sigma$ detection limits from 
Method 3 are shown, converted to planet mass
using models from \citet{bur}.  Planetary orbits around GJ 896 A
would be destabilized beyond about 12 AU by the companion star
GJ 896 B.  In order from bottom to top, the curves give
the 90th, 80th, 70th, 60th, 50th, 40th, 30th, 20th, 
10th, and 0th percentile sensitivity at each radius.}
\label{fig:percen1}
\end{figure*}

\begin{figure*}
\plottwo{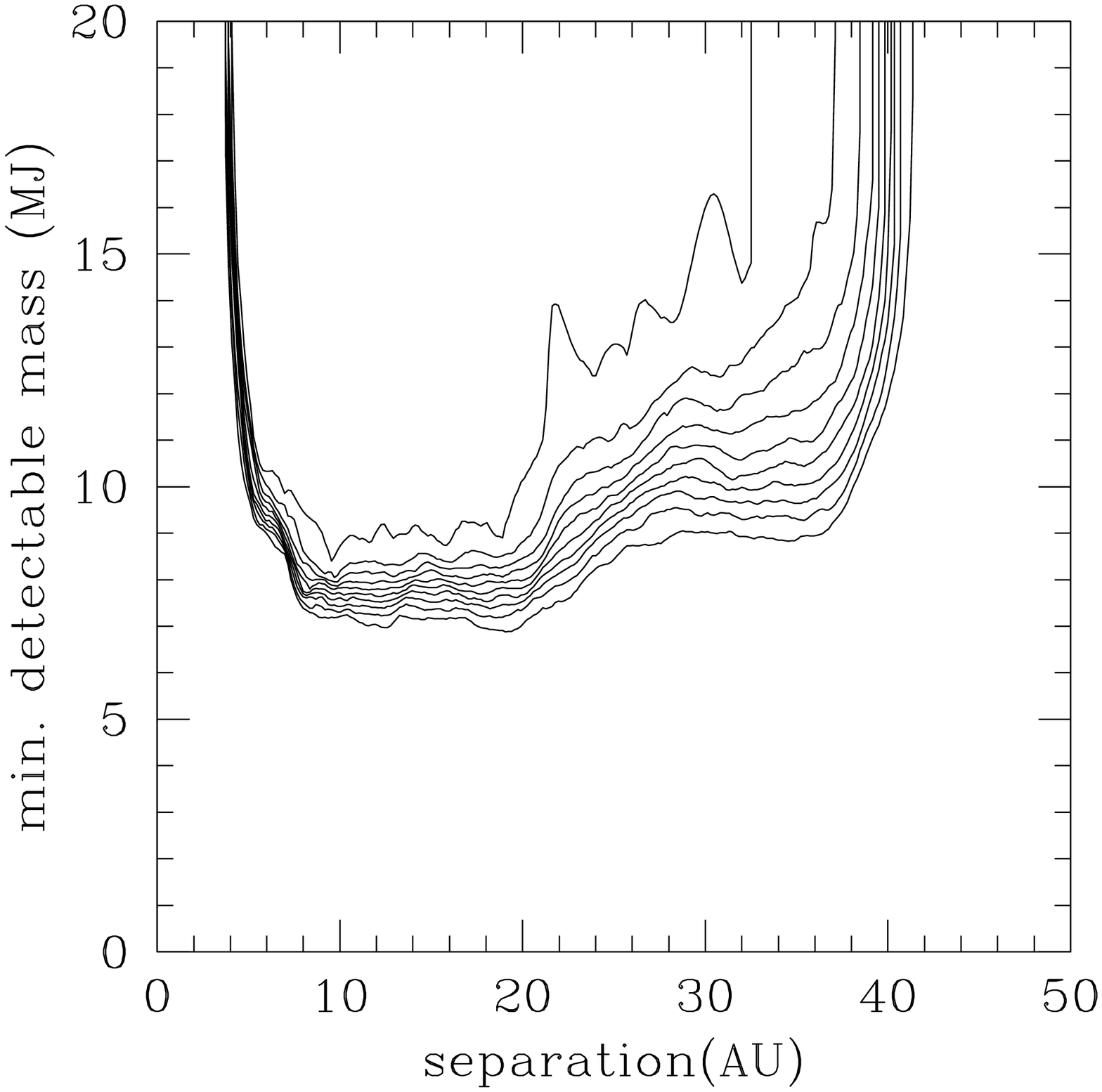}{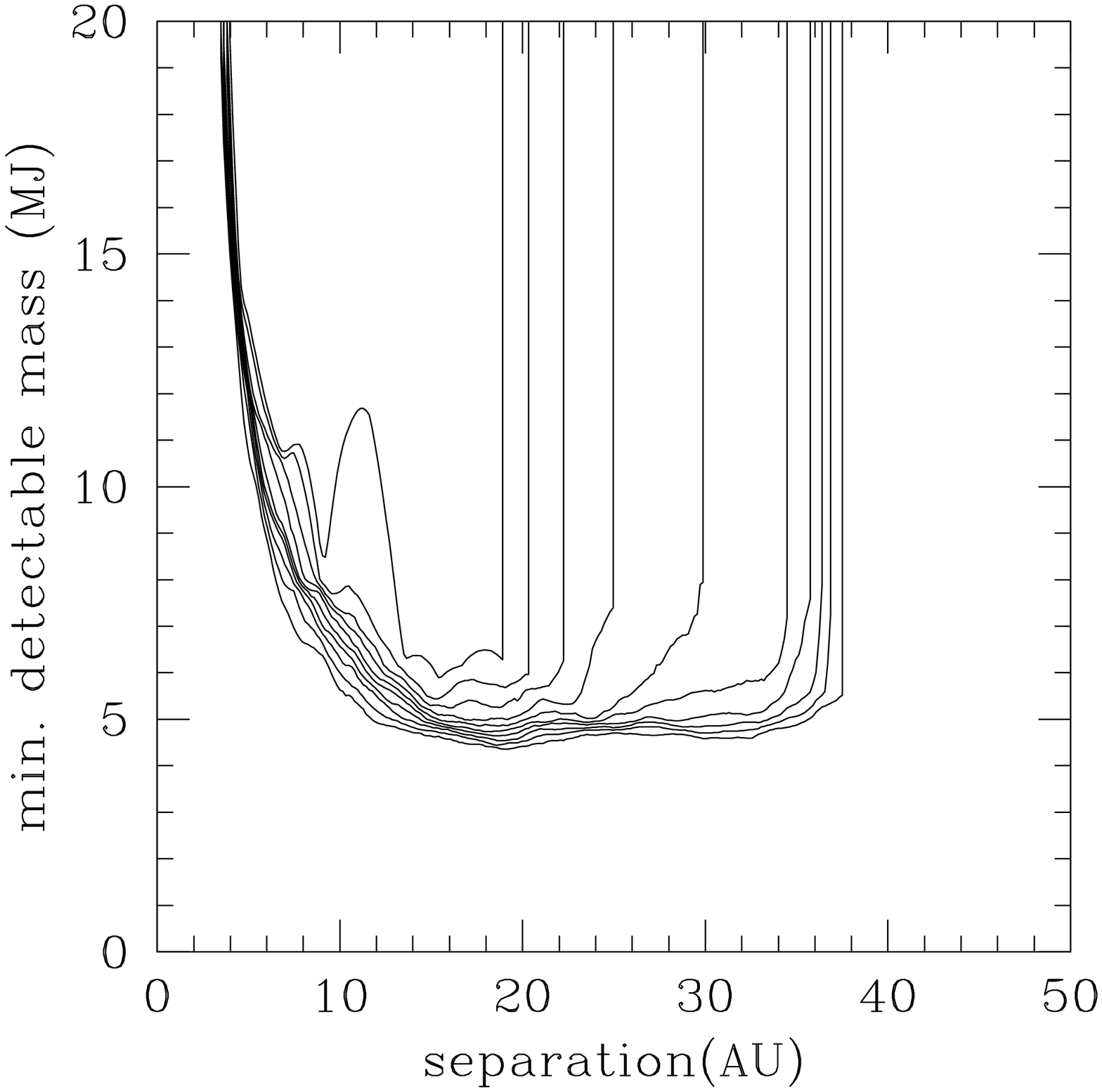}
\caption[Sensitivity Plots for 61 Cyg B ($M$ band), and $\epsilon$ Eri]
{Minimum detectable
planet mass vs. projected separation in AU for 
61 Cyg B ($M$ band data; left), and $\epsilon$ Eri (right).  
10$\sigma$ detection
limits from Method 3 are shown, converted to planet mass
using models from \citet{bur}. In order from bottom to top, the curves give
the 90th, 80th, 70th, 60th, 50th, 40th, 30th, 20th, 
10th, and 0th percentile sensitivity at each radius.}
\label{fig:percen2}
\end{figure*}

For use in the Monte Carlo simulations described in \citet{modeling},
we have converted our sensitivity maps into
plots of sensitivity vs. projected radius from each star.
As can be seen from Figures \ref{fig:GJ896cons} through \ref{fig:61CygBcons},
however, our sensitivity varied widely with position angle
around the star.  To quantify this, we calculated ten different
sensitivity values at each radius, giving the percentiles in sensitivity
from 0th to 90th percentile in 10\% increments.  Thus, e.g., the 0th
percentile at 2 arcsec is the very worst sensitivity obtained
anywhere on the 2 arcsec-radius ring surrounding the star, while the 50th
percentile gives the median sensitivity at that radius.
In Figures \ref{fig:percen1}  and \ref{fig:percen2},
we give example plots for GJ 896 A,
GJ 117, 61 Cyg B ($M$ band), and $\epsilon$ Eri, with
the sensitivities converted to minimum detectable
planet mass in \mjup~using models from \citet{bur}, plotted 
against projected separation in AU.  Plots of this type
for all the stars in our survey, as well as the tabular
data from which they were constructed, can be downloaded
from \url{http://www.hopewriter.com/Astronomyfiles/Data/SurveyPaper/}.

\subsection{Source Detection} \label{sec:sourcedet}

While our final sensitivity maps are constructed using only
Method 3, as described above, we use both Methods 2 and 3
for automated source detection.  The use of both methods
increases our likelihood of noticing faint sources at
the limit of detectability.  To search an image for sources
using either method, we query each pixel in turn to see
if a source is present at that location.  To make this
query, we first calculate the sensitivity statistic (Method 2 or Method 3)
over either a disk or an arc, just as described in
Section \ref{sec:estim}, except that a PSF-sized region
around the pixel being considered is not included, so that
if a real source is present, it will not bias the sensitivity
estimator.  Finally, either aperture photometry (Method 2)
or PSF-fitting (Method 3) is applied at the location of
the pixel itself, measuring the brightness of any source
that may be present there.  If the resulting brightness
is greater than the sensitivity statistic by a specified
threshold factor (i.e., 5 for a 5$\sigma$ detection), 
a preliminary detection is reported.

We would like to set the threshold as low as possible
without getting an unmanageable number of spurious
detections.  To this end, we divided each data set
into the first half of the images and the second
half, and created a stacked image from each half.
To be reported by our automated detection code,
a source had to appear at $4.5\sigma$ significance 
in the full stack, and at $3\sigma$ significance
on each half-stack, at a location consistent to
within 2 pixels.  This eliminated residual ghosts
and other artifacts, which would appear in different
locations on the two halves of the data due to
parallactic rotation.  Typically 10-20 spurious automated
detections were nonetheless reported around each star.

A real source could also be missed by the automatic
algorithm but noticed manually.  For example, due
to parallactic rotation, a location might 
have valid data only for the first half of the data
sequence, rendering an automated detection of a
real source there impossible.  Every automated detection,
as well as candidate sources noticed only by eye, was
carefully examined manually.  Criteria applied included
correct FWHM and symmetry, consistency in position and brightness
from one half-stack to the other, and inability to be 
explained away as an artifact of ghosts, diffraction rays, etc.
If necessary, data stacks were split into quarters or
even finer divisions to verify sources where only a
fraction of the images provided useful data.  These manual
investigations were very labor-intensive, especially since
the master images and half-stacks from several different
processing methods (see Section \ref{sec:datproc}) had to be
examined for each star.  Every
source that passed this final manual analysis was found
to correspond to a real astronomical object.  There were
no false positives.

\subsection{Blind Sensitivity Tests} \label{sec:blind}

The final demonstration of the validity of a sensitivity
estimator is a blind sensitivity test, in which fake planets
are inserted into the raw data and then recovered by an experimenter
(or automated process) without a-priori knowledge of their 
positions or their number.  Such a blind test is the surest
way to evaluate any sensitivity estimator and establish
the relationship between nominal significance (i.e. 3$\sigma$,
5$\sigma$, etc.) and the true completeness level of the survey.
This should be standard procedure for all planet imaging surveys.

We inserted simulated planets
at random locations in the raw data for selected stars.  
The flux of each simulated planet was scaled to 5, 7, or 10$\sigma$ 
significance based on the master 
sensitivity map (see Section \ref{sec:realsens}) for that star.
The PSFs for the planets were taken from the short exposure,
unsaturated images of the parent star, mentioned above in Section \ref{sec:obs}.
The raw data with fake planets inserted was then processed 
exactly as for the real, unmodified
science data for that star, and planets were sought in the 
fully processed images by the same combination of manual and 
automatic methods used for the real images.

The final result of each test was that every inserted planet
was classified as `Confirmed', `Noticed', or `Unnoticed'.
`Confirmed' means the source was confidently detected, with
no significant doubt of its being a real object.  
`Noticed' means the source was flagged by our automatic detection algorithm,
or noticed manually as a possible real object, but could not 
be confirmed beyond reasonable doubt.  Many spurious sources are 
`Noticed' whereas the false-positive rate for `Confirmed' detections 
is extremely low, with none for any of the data
sets discussed here.  `Unnoticed' means a fake planet was not automatically
flagged or noticed manually.

\begin{figure*}
\plotone{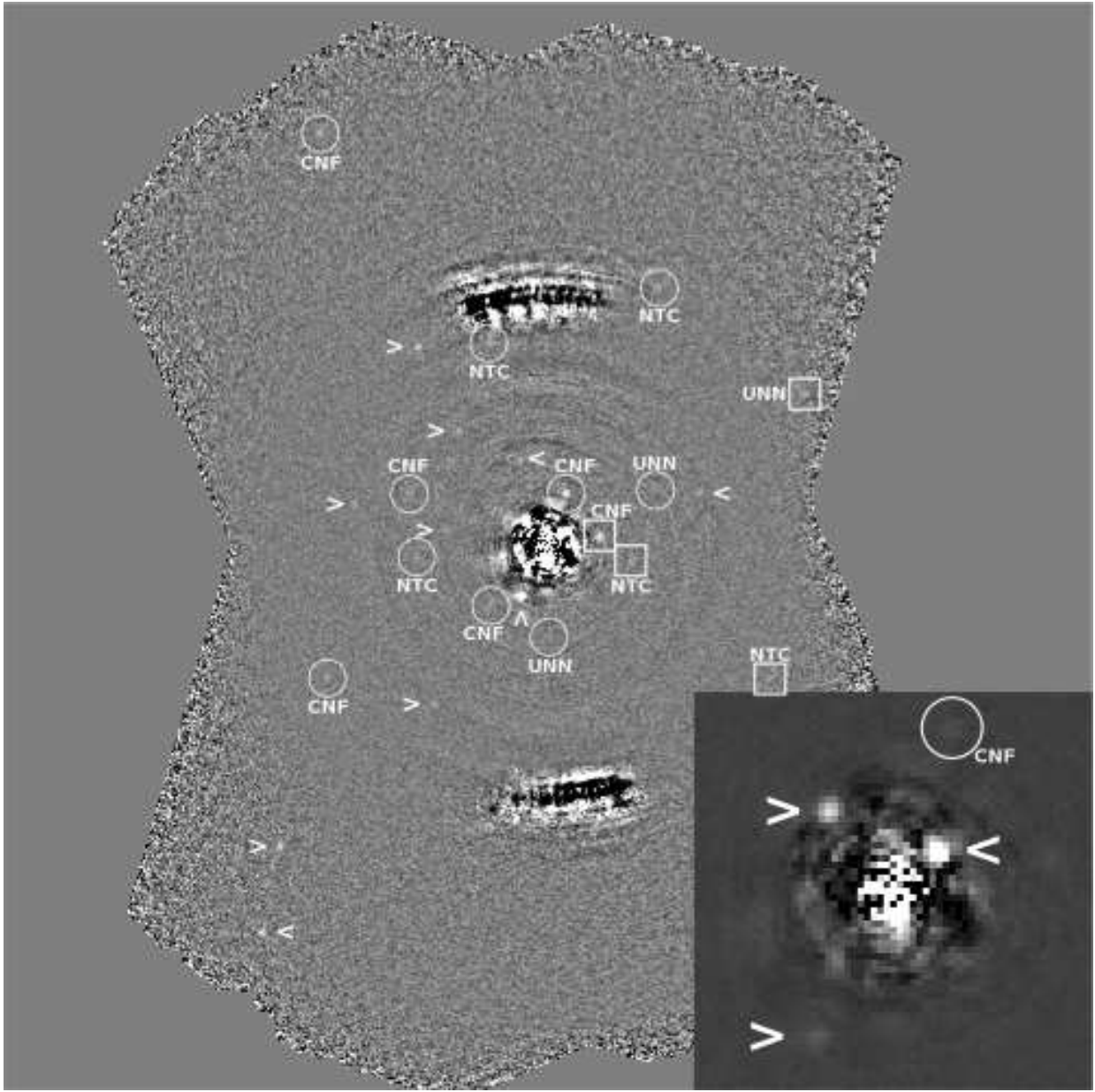}
\caption[Image from Blind Sensitivity Test on HD 29391]{Fully
processed `e' method master image from the blind sensitivity
test on HD 29391.  In this data set there
are 12 planets of 10$\sigma$ significance (indicated
by arrows), 10 at 7$\sigma$ (circled),
and 5 at 5$\sigma$ (boxed).  One 5$\sigma$ planet is hidden by the
inset.  Each planet was either confirmed (CNF), noticed (NTC),
or unnoticed (UNN) in the blind test.  All 10$\sigma$ planets
were confirmed.  The inset, 3 arcsec square, shows the inner
part of the image magnified 3$\times$ and with display range
increased 10$\times$ relative to the main image.  The main 
image is 24 arcsec square.  Two planets are marked in both
the main image and the inset.}  
\label{fig:Hurin}
\end{figure*}

Tables \ref{tab:GJ450sim} through \ref{tab:BD3686sim} give
the results of these simulations, showing how the detectable
planet masses vary with the distance and age of the stars,
and with data quality.  Note that simulated planets
with masses ranging down to 3 \mjup~and below were confirmed, the lowest
mass planet confirmed being one of 2.36 \mjup~in the GJ 117 simulation.
Figure \ref{fig:Hurin} shows an image from our blind sensitivity
test on HD 29391, with the simulated planets marked.  The
random positions of the planets, unknown by the experimenter attempting to
detect the them, are an important aspect of our tests.

\begin{deluxetable}{rrrrc}
\tablecaption{GJ 450 fake planet experiment.\label{tab:GJ450sim}}
\tablewidth{0pt}
\tablehead{\colhead{Sep} & & \colhead{Mass} & \colhead{Detection} & \\
\colhead{(arcsec)} & \colhead{$L'$ Mag} & \colhead{(\mjup)} & \colhead{Significance} & \colhead{Status}}
\startdata
0.51 & 12.53 & $>$20 & 10.00$\sigma$ & Confirmed \\
0.56 & 13.32 & $>$20 & 10.00$\sigma$ & Confirmed \\
0.95 & 15.35 & 11.26 & 10.00$\sigma$ & Confirmed \\
1.14 & 15.60 & 10.54 & 10.00$\sigma$ & Confirmed \\
1.27 & 15.96 & 9.51 & 10.00$\sigma$ & Confirmed \\
1.58 & 16.06 & 9.21 & 10.00$\sigma$ & Confirmed \\
1.90 & 16.51 & 7.93 & 10.00$\sigma$ & Confirmed \\
2.50 & 16.59 & 7.73 & 10.00$\sigma$ & Confirmed \\
2.69 & 16.57 & 7.78 & 10.00$\sigma$ & Confirmed \\
2.91 & 16.38 & 8.29 & 10.00$\sigma$ & Confirmed \\
2.98 & 16.60 & 7.70 & 10.00$\sigma$ & Confirmed \\
3.71 & 16.51 & 7.93 & 10.00$\sigma$ & Confirmed \\
3.90 & 16.59 & 7.73 & 10.00$\sigma$ & Confirmed \\
3.93 & 16.62 & 7.65 & 10.00$\sigma$ & Confirmed \\
5.02 & 16.49 & 7.98 & 10.00$\sigma$ & Confirmed \\
6.52 & 16.43 & 8.15 & 10.00$\sigma$ & Confirmed \\
6.53 & 16.27 & 8.61 & 10.00$\sigma$ & Confirmed \\
\enddata
\tablecomments{All of the input planets were
confirmed.  Planet magnitude to mass conversion carried out by
interpolation based on theoretical spectra from \citet{bur}, using
our adopted distance and age for this star (8.1 pc, 1.0 Gyr).}
\end{deluxetable}

\begin{deluxetable}{rrrrc}
\tablecaption{HD 29391 fake planet experiment.\label{tab:HD29391sim}}
\tablewidth{0pt}
\tablehead{\colhead{Sep} & & \colhead{Mass} & \colhead{Detection} & \\
\colhead{(arcsec)} & \colhead{$L'$ Band Mag} & \colhead{(\mjup)} & \colhead{Significance} & \colhead{Status}}
\startdata
0.42 & 11.59 & $>$20 & 10.00$\sigma$ & Confirmed \\
0.76 & 12.56 & 16.85 & 10.00$\sigma$ & Confirmed \\
1.23 & 15.35 & 4.97 & 10.00$\sigma$ & Confirmed \\
2.06 & 15.90 & 3.92 & 10.00$\sigma$ & Confirmed \\
2.27 & 16.10 & 3.63 & 10.00$\sigma$ & Confirmed \\
3.26 & 14.58 & 6.95 & 10.00$\sigma$ & Confirmed \\
3.60 & 15.77 & 4.15 & 10.00$\sigma$ & Confirmed \\
4.29 & 15.48 & 4.72 & 10.00$\sigma$ & Confirmed \\
4.41 & 16.22 & 3.46 & 10.00$\sigma$ & Confirmed \\
5.31 & 16.21 & 3.47 & 10.00$\sigma$ & Confirmed \\
8.92 & 16.15 & 3.56 & 10.00$\sigma$ & Confirmed \\
10.69 & 16.15 & 3.56 & 10.00$\sigma$ & Confirmed \\
1.25 & 15.17 & 5.40 & 7.00$\sigma$ & Confirmed \\
1.86 & 16.32 & 3.31 & 7.00$\sigma$ & Confirmed \\
2.00 & 16.47 & 3.09 & 7.00$\sigma$ & Unnoticed \\
2.69 & 16.54 & 2.99 & 7.00$\sigma$ & Unnoticed \\
2.92 & 16.61 & 2.93 & 7.00$\sigma$ & Noticed \\
3.29 & 16.47 & 3.09 & 7.00$\sigma$ & Confirmed \\
4.69 & 15.83 & 4.03 & 7.00$\sigma$ & Noticed \\
5.72 & 16.38 & 3.22 & 7.00$\sigma$ & Confirmed \\
6.28 & 15.97 & 3.82 & 7.00$\sigma$ & Noticed \\
10.53 & 15.94 & 3.86 & 7.00$\sigma$ & Confirmed \\
1.19 & 15.39 & 4.89 & 5.00$\sigma$ & Confirmed \\
1.93 & 16.77 & 2.78 & 5.00$\sigma$ & Noticed \\
5.76 & 16.57 & 2.97 & 5.00$\sigma$ & Noticed \\
6.68 & 16.25 & 3.41 & 5.00$\sigma$ & Unnoticed \\
7.70 & 16.18 & 3.51 & 5.00$\sigma$ & Unnoticed \\
\enddata
\tablecomments{Planets confirmed: 12/12 at 10$\sigma$; 5/10 at 7$\sigma$;
1/5 at 5$\sigma$.  Planets noticed: 12/12 at 10$\sigma$; 8/10 at 7$\sigma$;
3/5 at 5$\sigma$. Planet magnitude to mass conversion carried out by
interpolation based on theoretical spectra from \citet{bur}, using
our adopted distance and age for this star (14.71 pc, 0.1 Gyr).}
\end{deluxetable}

\begin{deluxetable}{rrrcc}
\tablecaption{GJ 117 fake planet experiment.\label{tab:GJ117sim}}
\tablewidth{0pt}
\tablehead{\colhead{Sep} & & \colhead{Mass} & \colhead{Detection} & \\
\colhead{(arcsec)} & \colhead{$L'$ Band Mag} & \colhead{(\mjup)} & \colhead{Significance} & \colhead{Status}}
\startdata
0.67 & 10.41 & $>$20.0 & 10.00$\sigma$ & Confirmed \\
0.94 & 11.54 & 15.42 & 10.00$\sigma$ & Confirmed \\
1.10 & 12.05 & 12.21 & 10.00$\sigma$ & Confirmed \\
2.11 & 15.01 & 3.42 & 10.00$\sigma$ & Confirmed \\
2.17 & 14.78 & 3.75 & 10.00$\sigma$ & Confirmed \\
3.31 & 14.93 & 3.53 & 10.00$\sigma$ & Confirmed \\
3.77 & 15.20 & 3.14 & 10.00$\sigma$ & Confirmed \\
6.40 & 14.72 & 3.84 & 10.00$\sigma$ & Confirmed \\
6.42 & 15.26 & 3.05 & 10.00$\sigma$ & Confirmed \\
8.60 & 15.06 & 3.35 & 10.00$\sigma$ & Confirmed \\
9.88 & 14.56 & 4.09 & 10.00$\sigma$ & Confirmed \\
1.14 & 12.54 & 9.77 & 7.00$\sigma$ & Confirmed \\
3.08 & 15.44 & 2.87 & 7.00$\sigma$ & Noticed \\
5.06 & 15.35 & 2.96 & 7.00$\sigma$ & Confirmed \\
6.37 & 14.67 & 3.91 & 7.00$\sigma$ & Noticed \\
7.04 & 14.66 & 3.93 & 7.00$\sigma$ & Noticed \\
7.88 & 15.27 & 3.05 & 7.00$\sigma$ & Noticed \\
1.04 & 12.31 & 10.83 & 5.00$\sigma$ & Confirmed \\
1.75 & 15.12 & 3.26 & 5.00$\sigma$ & Unnoticed \\
2.89 & 15.96 & 2.40 & 5.00$\sigma$ & Unnoticed \\
3.30 & 16.16 & 2.21 & 5.00$\sigma$ & Unnoticed \\
5.08 & 16.00 & 2.36 & 5.00$\sigma$ & Confirmed \\
7.80 & 15.32 & 2.98 & 5.00$\sigma$ & Noticed \\
8.03 & 15.65 & 2.68 & 5.00$\sigma$ & Unnoticed \\
10.21 & 15.30 & 3.00 & 5.00$\sigma$ & Noticed \\
\enddata
\tablecomments{Planets confirmed: 11/11 at 10$\sigma$; 2/6 at 7$\sigma$;
2/8 at 5$\sigma$.  Planets noticed: 11/11 at 10$\sigma$; 6/6 at 7$\sigma$;
4/8 at 5$\sigma$. Planet magnitude to mass conversion carried out by
interpolation based on theoretical spectra from \citet{bur}, using
our adopted distance and age for this star (8.31 pc, 0.1 Gyr).
Note that a fake planet with a mass of only 2.36 \mjup~was confirmed.}
\end{deluxetable}

\begin{deluxetable}{rrrrc}
\tablecaption{GJ 355 fake planet experiment.\label{tab:GJ355sim}}
\tablewidth{0pt}
\tablehead{\colhead{Sep} & & \colhead{Mass} & \colhead{Detection} & \\
\colhead{(arcsec)} & \colhead{$L'$ Band Mag} & \colhead{(\mjup)} & \colhead{Significance} & \colhead{Status}}
\startdata
0.37 & 9.46 & $>$20.0 & 10.00$\sigma$ & Confirmed \\
0.43 & 9.66 & $>$20.0 & 10.00$\sigma$ & Confirmed \\
0.94 & 13.72 & 13.10 & 10.00$\sigma$ & Confirmed \\
1.67 & 15.61 & 5.74 & 10.00$\sigma$ & Confirmed \\
1.74 & 15.66 & 5.63 & 10.00$\sigma$ & Confirmed \\
1.85 & 15.74 & 5.43 & 10.00$\sigma$ & Confirmed \\
2.05 & 15.63 & 5.70 & 10.00$\sigma$ & Confirmed \\
2.37 & 15.87 & 5.11 & 10.00$\sigma$ & Noticed \\
3.08 & 15.60 & 5.78 & 10.00$\sigma$ & Confirmed \\
3.30 & 15.92 & 5.00 & 10.00$\sigma$ & Confirmed \\
3.44 & 15.73 & 5.46 & 10.00$\sigma$ & Confirmed \\
4.26 & 16.02 & 4.80 & 10.00$\sigma$ & Confirmed \\
5.55 & 15.87 & 5.12 & 10.00$\sigma$ & Confirmed \\
8.09 & 15.55 & 5.89 & 10.00$\sigma$ & Confirmed \\
8.70 & 15.34 & 6.46 & 10.00$\sigma$ & Confirmed \\
1.57 & 15.95 & 4.93 & 7.00$\sigma$ & Noticed \\
2.83 & 16.24 & 4.37 & 7.00$\sigma$ & Noticed \\
3.68 & 16.04 & 4.77 & 7.00$\sigma$ & Confirmed \\
4.34 & 16.01 & 4.82 & 7.00$\sigma$ & Confirmed \\
4.68 & 16.33 & 4.19 & 7.00$\sigma$ & Noticed \\
6.99 & 15.95 & 4.94 & 7.00$\sigma$ & Confirmed \\
1.92 & 16.58 & 3.78 & 5.00$\sigma$ & Unnoticed \\
3.24 & 16.52 & 3.87 & 5.00$\sigma$ & Unnoticed \\
5.61 & 15.93 & 4.99 & 5.00$\sigma$ & Noticed \\
5.99 & 15.86 & 5.16 & 5.00$\sigma$ & Noticed \\
7.17 & 15.94 & 4.97 & 5.00$\sigma$ & Noticed \\
10.07 & 16.31 & 4.23 & 5.00$\sigma$ & Confirmed \\
\enddata
\tablecomments{Planets confirmed: 14/15 at 10$\sigma$; 3/6 at 7$\sigma$;
1/6 at 5$\sigma$.  Planets noticed: 15/15 at 10$\sigma$; 6/6 at 7$\sigma$;
4/6 at 5$\sigma$. Planet magnitude to mass conversion carried out by
interpolation based on theoretical spectra from \citet{bur}, using
our adopted distance and age for this star (19.23 pc, 0.1 Gyr).}
\end{deluxetable}

\begin{deluxetable}{rrrrc}
\tablecaption{BD+48 3686 fake planet experiment.\label{tab:BD3686sim}}
\tablewidth{0pt}
\tablehead{\colhead{Sep} & & \colhead{Mass} & \colhead{Detection} & \\
\colhead{(arcsec)} & \colhead{$L'$ Band Mag} & \colhead{(\mjup)} & \colhead{Significance} & \colhead{Status}}
\startdata
0.23 & 8.03 & $>$20.0 & 10.00$\sigma$ & Confirmed \\
0.97 & 14.65 & 13.89 & 10.00$\sigma$ & Noticed \\
1.33 & 15.19 & 10.47 & 10.00$\sigma$ & Confirmed \\
2.05 & 15.51 & 9.05 & 10.00$\sigma$ & Confirmed \\
4.33 & 15.57 & 8.85 & 10.00$\sigma$ & Confirmed \\
5.08 & 15.70 & 8.41 & 10.00$\sigma$ & Confirmed \\
6.13 & 15.52 & 9.04 & 10.00$\sigma$ & Confirmed \\
6.34 & 14.70 & 13.53 & 10.00$\sigma$ & Confirmed \\
8.41 & 15.38 & 9.60 & 10.00$\sigma$ & Confirmed \\
9.73 & 15.46 & 9.26 & 10.00$\sigma$ & Confirmed \\
1.46 & 15.62 & 8.67 & 7.00$\sigma$ & Confirmed \\
2.55 & 15.86 & 7.87 & 7.00$\sigma$ & Noticed \\
3.76 & 16.15 & 7.05 & 7.00$\sigma$ & Unnoticed \\
5.25 & 15.72 & 8.32 & 7.00$\sigma$ & Confirmed \\
5.73 & 15.66 & 8.53 & 7.00$\sigma$ & Unnoticed \\
10.43 & 15.41 & 9.50 & 7.00$\sigma$ & Confirmed \\
1.08 & 15.63 & 8.66 & 5.00$\sigma$ & Noticed \\
3.04 & 16.39 & 6.45 & 5.00$\sigma$ & Unnoticed \\
3.34 & 16.29 & 6.70 & 5.00$\sigma$ & Unnoticed \\
5.69 & 16.40 & 6.42 & 5.00$\sigma$ & Noticed \\
9.19 & 16.17 & 7.00 & 5.00$\sigma$ & Unnoticed \\
10.22 & 15.97 & 7.56 & 5.00$\sigma$ & Noticed \\
\enddata
\tablecomments{Planets confirmed: 9/10 at 10$\sigma$; 3/6 at 7$\sigma$;
0/6 at 5$\sigma$.  Planets noticed: 10/10 at 10$\sigma$; 4/6 at 7$\sigma$;
3/6 at 5$\sigma$. Planet magnitude to mass conversion carried out by
interpolation based on theoretical spectra from \citet{bur}, using
our adopted distance and age for this star (23.6 pc, 0.15 Gyr).}
\end{deluxetable}

The total statistics from all 5 blind tests are that
63 of 65 planets were confirmed at 10$\sigma$, 13 of 28
at 7$\sigma$, and 4 of 25 at 5$\sigma$.  In percentages
we have 97\% completeness at 10$\sigma$, 46\% completeness
at 7$\sigma$, and 16\% completeness at 5$\sigma$.  

Note the very low completeness at 5$\sigma$, which many
past surveys have taken as a realistic detection limit.
Though sensitivity estimators (and therefore
the exact meaning of 5$\sigma$) differ, ours was quite
conservative.  The low completeness we find at 5$\sigma$ should
serve as a warning to future workers in this field, and an
encouragement to establish a definitive significance-completeness
relation through blind sensitivity tests as 
we have done.  Many more planets were noticed
than were confirmed: for noticed planets, the rates
are 100\% at 10$\sigma$, 86\% at 7$\sigma$, and 56\% at 5$\sigma$.
However, very many false positives were also noticed, so
sources that are merely noticed but not confirmed do not
represent usable detections.  No false positives were
confirmed in any of our blind tests.

There are several reasons for our low
completeness rate at $5\sigma$.  First, some flux is lost from
faint sources in our processing, as described above, so
that sources input at 5$\sigma$ significance are reduced
to a real significance of typically 4$\sigma$ in the final
image.  Second, since our images contain speckles, ghosts, 
diffraction rays, and pattern noise, the noise is not 
gaussian but rather has a long tail toward improbable, bright events --
a normal circumstance in AO images that has been carefully
described by \citet{fgspeckle} and \citet{mspeckle}.  
Third, the area of each final image is over $10^5$ times
the size of a PSF, so the distribution of possible 
spurious planet images arising from noise is sampled at least
$10^5$ times for each final image in our survey.  Followup
observations of suspected sources are costly in terms of
telescope time, so a detection strategy with a low false-positive
rate is important.  

While background noise originating from photon statistics
in astronomical images is gaussian, speckle noise in AO-corrected 
images close to bright stars has been shown to follow
a longer tailed, approximately rician distribution \citep{mspeckle,fgspeckle}.
In fact, \citet{mspeckle} have shown that to obtain an
acceptably low false positive rate, detection thresholds
must be set as high as $12 \sigma$ in the presence of severe
speckle noise.  They assume a detection strategy based on the
single-pixel RMS standard deviation (e.g. \citet{biller1,chauvin})
rather than sensitivity estimation methods like ours or
that of \citet{GDPS}.  Even so, given their findings it may seem
surprising not that we had low completeness at $5 \sigma$, but
that we \textit{were} able to detect some $5 \sigma$ sources while
also maintaining a very low false-positive rate.

Part of the explanation for this is the speckle-supression
produced by ADI: \citet{mspeckle} found that ADI could be
so powerful that it nearly restored gaussian statistics
to an image, allowing the viable detection threshold to drop
from $12 \sigma$ to lower than $6 \sigma$.  Our implementation of
ADI may not be as effective as that of \citet{mspeckle}, but it
did substantially improve our image statistics.  This is demonstrated
by the fact that our blind sensitivity tests did not show any clear bias
against detection of low-significance planets close to the star.
However, some of our ability to confirm low-significance planets
is simply due to our painstaking detection strategy.  Noise-bursts
at 10 or 12 $\sigma$ may occur in the speckle-dominated regions of
AO images, but splitting the data in half, examining master images
created using different processing methods, and other time-intensive
analyses can powerfully sort out the real from the unreal -- even, in
some cases, when the spurious sources are substantially brighter.

Accurate estimation of the sensitivity of AO images is a complex task,
worthy, perhaps, of more attention than has been paid it in the AO
planet-search literature up to this point.  Between this work, \citet{GDPS},
\citet{biller1} and \citet{kasper}, and others, several different
sensitivity estimators have been used, which may produce substantially
different results. Statistical noise distributions
can vary widely even on a single image \citep{mspeckle}, and certainly
exhibit further variability from instrument to instrument and telescope
to telescope.  A blind sensitivity test such as we have carried out is
an excellent way to determine the true sensitivity of a set of observations.
Completeness vs. significance relations established by such blind
sensitivity tests may represent the only real option for `apples-to-apples'
comparisons of the sensitivity obtained with different instruments on different
telescopes -- and such comparisons may be quite important for selecting
optimal observing strategies as we move forward to the next generation
of surveys to detect extrasolar planets.

\section{Detections of Faint Real Objects}\label{sec:faintreal}

\subsection{Overview of Detected Companions}\label{sec:back}
In all, thirteen faint sources were confirmed as real.  
Table \ref{tab:sources} presents our astrometry and photometry
for each detected companion.

\begin{deluxetable}{lrrrrc}
\tablewidth{0pc}
\tablecolumns{6}
\tablecaption{Confirmed Sources in Our Survey\label{tab:sources}}
\tablehead{\colhead{Star} & \colhead{Det.} & \colhead{$L'$} & \colhead{Sep} & & \colhead{Date}\\
\colhead{Name} & \colhead{Sig.} & \colhead{Mag} & \colhead{(arcsec)} & \colhead{PA} & \colhead{(yyyy/mm/dd)}}
\startdata
GJ 354.1A & 4.93$\sigma$ & $16.48 \pm 0.20$ & 4.93 & $187.3^{\circ}$ & 2006/04/12\\ 
GJ 564 & 175.68$\sigma$ & $10.80 \pm 0.20$ & 2.60 & $103.0^{\circ}$ & 2006/04/13\\ 
GJ 3876 & 246.38$\sigma$ & $10.88 \pm 0.20$ & 1.85 & $118.6^{\circ}$ & 2006/04/13\\ 
GJ 3860 & 19.21$\sigma$ & $14.90 \pm 0.20$ & 9.68 & $144.4^{\circ}$ & 2006/06/09\\ 
61 Cyg A & \nodata & $12.43 \pm 0.20$ & 11.24 & $227.5^{\circ}$ & 2006/06/09\\ 
61 Cyg A & 32.82$\sigma$ & $13.05 \pm 0.20$ & 7.78 & $83.2^{\circ}$ & 2006/06/09\\ 
61 Cyg B & \nodata & $14.04 \pm 0.20$ & 9.85 & $145.4^{\circ}$ & 2006/06/10\\ 
BD+60 1417 & 11.91$\sigma$ & $15.71 \pm 0.20$ & 1.93 & $301.4^{\circ}$ & 2006/06/10\\ 
GJ 684 A & 7.23$\sigma$ & $15.00 \pm 0.20$ & 3.01 & $358.5^{\circ}$ & 2006/06/11\\
GJ 860 A & \nodata & $15.76 \pm 0.20$ & 7.24 & $0.25^{\circ}$ & 2006/06/12\\
BD+20 1790 & 31.51$\sigma$ & $14.60 \pm 0.20$ & 8.73 & $74.1^{\circ}$ & 2007/01/04 \\ 
BD+20 1790 & \nodata & $15.16 \pm 0.20$ & 6.42 & $336.4^{\circ}$ & 2007/01/04 \\ 
HD 96064A & 43.18$\sigma$ & $13.72 \pm 0.20$ & 5.57 & $212.8^{\circ}$ & 2007/01/04\\ 
\enddata
\tablecomments{The detection significance column gives the highest significance
with which the source was automatically detected on any image with any method.
Blanks in this column imply sources that were detected only manually.
Uncertainties on the astrometry are about 0.05 arcsec or less; note that
the position angle values of close-in companions are thus more
uncertain than those of distant ones.  Some of the photometry
may be more accurate than the 0.2 mag uncertainties we have conservatively
quoted.  The photometry of GJ 564 is probably too faint because
the aperture correction will not have been accurate for this close
binary.}
\end{deluxetable}

Of these 13 faint companions,
one is a newly discovered low mass star orbiting GJ 3876
(see Section \ref{sec:discovery}),
one is a previously known binary brown dwarf companion
to GJ 564 \citep{potter}, and the other eleven are background stars.  
Note that \citet{GDPS}, operating in the
$H$ band regime, found more than 300 background stars.
Due to the red IR colors of planets, a long wavelength
survey such as ours can obtain good sensitivity
to planets while remaining blind to all but the brightest stars,
so that less telescope time is needed to follow up
candidate objects.  Also, a background star masquerading
as a planet at $L'$ can often be detected in a short integration
at shorter wavelengths, showing that the
object is far too blue in IR color to be a planet.  
We have applied this strategy by taking $K_S$ band images
of the brighter of the two companions of BD+20 1790, 
and the faint companions near HD 96064, BD+60 1417, 
and GJ 3860, in all cases obtaining bright $K_S$ band
fluxes that indicate the objects have stellar $K_S - L'$
color, rather than the very red $K_S - L'$ colors
expected for planets.  Such color measurements can often
rule out a planet candidate immediately, in contrast to
the waiting period required for proper motion confirmation.

For planets near our detection limit, expected $K_S - L'$ colors
are generally so red that a $K_S$ detection effectively
rules out the candidate.  However, for brighter candidates,
the case is not always so clear, as these would correspond
to hotter planets with less red colors.  We note that the 
sources we detected around HD 96064 A and BD+60 1417
were independently detected in the \citet{GDPS} survey,
and confirmed to be background objects based on proper
motions.  The HD 96064 A source looks double in our data,
and was confirmed to be so by \citet{GDPS}.  There are
only two sources we classified as background objects based
on color alone: the one near GJ 3860, and the brighter of the two near
BD+20 1790.  As these were well above our detection limit,
we consider whether the expected colors are red enough
relative to the measured ones to rule out a planetary
interpretation.

For the GJ 3860 companion, we obtained a brightness of
$L' = 14.90 \pm 0.10$.  According to the models of \citet{bur},
at the age and distance we adopted for this system, $L' = 14.9$
corresponds to an 11.3 \mjup~planet with $K_S - L' = 2.14$.
Using the models of \citet{bar} instead yields a 9.1 \mjup~planet
with $K_S - L = 3.9$. Our measured color for the object was 
$K_S - L' = 0.26 \pm 0.22$, consistent with a background star 
of any spectral type between F and late M (see Tables 7.6-7.8
in \citet{aaq}), but dramatically inconsistent with a planetary
interpretation based on either the \citet{bur} or the \citet{bar}
models.  The very different prediction from the two model sets
stems mainly from the different planet masses they imply for
the observed $L'$ magnitude: the models do not disagree so
widely on $K_S - L'$ color for a specific planet mass.
While the descrepancy does indicate considerable model
uncertainty, there is agreement that planets in the range
of mass and age applicable to this candidate are much 
redder in $K_S - L'$ color than stars. This is, of course, also 
our first-order expectation given the far lower effective 
temperatures of planets. The conclusion that the object we 
detected near GJ 3860 is a background star rather than a 
planet seems secure.

The case of BD+20 1790 is less clear-cut.  Our measured
$L'$ magnitude is $14.55 \pm 0.15$.  This implies a planet mass
of 16.75 \mjup~and a $K_S - L'$ color of 1.06 according to the
\citet{bur} models.  Our observed color is $K_S - L' = 0.36 \pm 0.20$.
While formally excluded, the planetary hypothesis does not
seem as untenable as for GJ 3860.  Using the \citet{bar} models
instead gives a 10.71 \mjup~planet with $K_S - L' = 2.71$, much
more comfortably excluded by the data.  The low galactic latitude
of BD+20 1790 (+16$^{\circ}$), combined with the presence of another
apparent companion (which was shown to be a background object
based on an archival HST image), suggests that there is a comparitively
rich star field behind BD+20 1790, and that the most plausible
interpretation of the brighter companion is, again, a
background star.  While interpretation as a planetary or brown
dwarf companion is perhaps not absolutely excluded, it is
much less likely a priori, and is inconsistent with the
observed color under both the model sets we have employed.

The companion of GJ 354.1 A is confirmed to be a background star
rather than a common proper motion companion based on an image by \citet{GJ3541A}.
The fainter of the two companions of BD+20 1790 is similarly
shown to be a background object by an archival HST
image.  The companions of 61 Cyg A and B are background objects
based on detections on POSS plates from 1991, when, due to
the 61 Cyg system's fast proper motion, the objects were much farther 
from the bright stars and therefore beyond the glare on the POSS images.
The companion of GJ 860 is confirmed to be a background star
based on previous detections on POSS plates from 1953, and
optical images of our own taken with the University of Arizona
1.5m Kuiper Telescope in 2005 (the latter simply prove the
object is too bright in the optical to be a planet).  The
POSS position match is imperfect, and our optical detection
is at low significance, but taken together they confirm
the object's nature.  The companion of GJ 684 is shown to
be a background star based on proper motion in followup
images we obtained using Clio in September 2008.

\begin{figure*}
\plotone{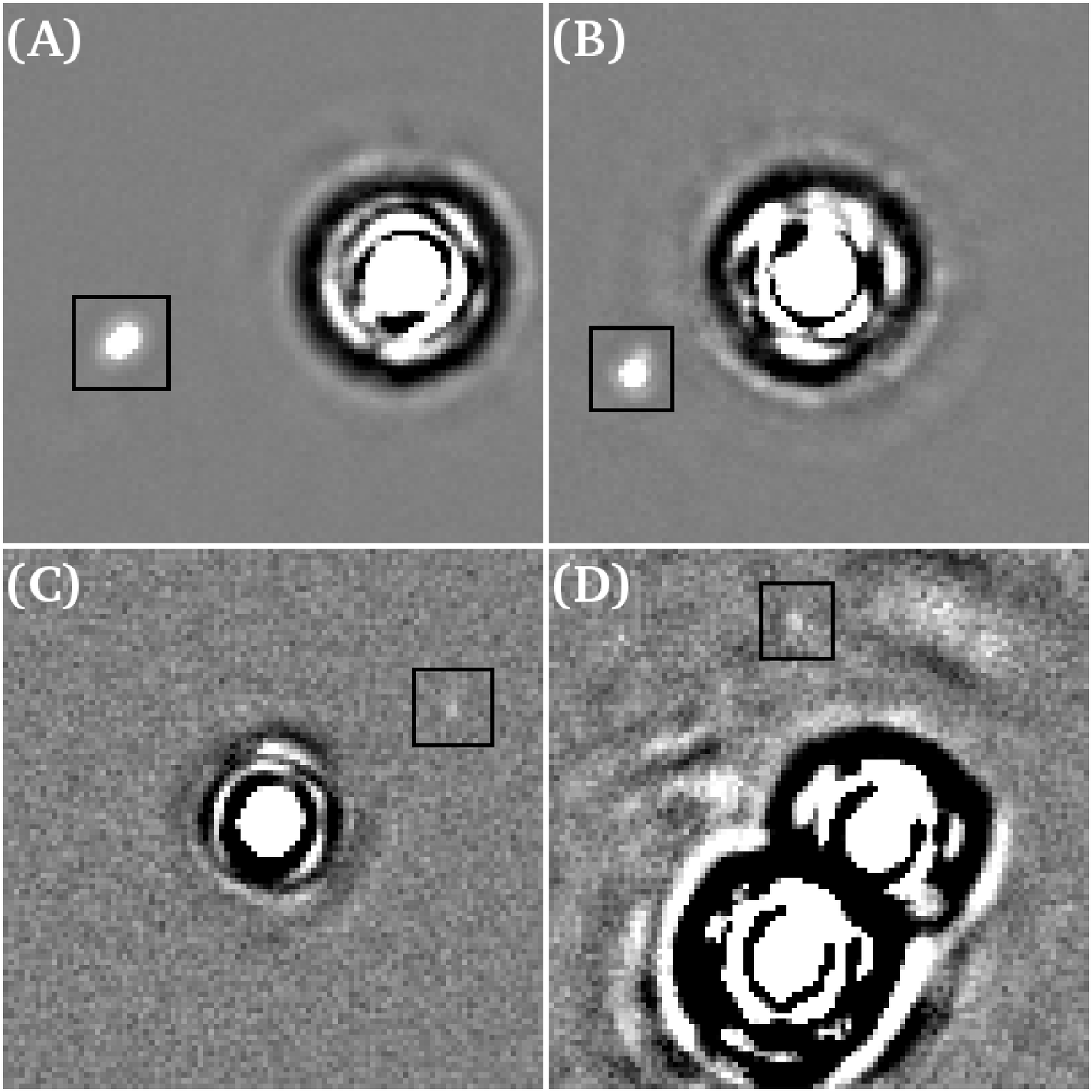}
\caption[Close-in Compansions Detected]{(A) $L'$ image
of GJ 564, showing the binary brown dwarf discovered by
\citet{potter}.  (B) $L'$ image of GJ 3876, showing
the low-mass stellar companion we discovered.
(C) $L'$ image of BD+60 1417, showing the faint
background star we detected.  (D) $L'$ image of binary
star GJ 684, showing the faint background star we
detected.  Each tile is 4.86 arcsec square;
the bottom tiles are contrast stretched 10$\times$ more than
the top ones to reveal the faint companions.} 
\label{fig:comp1}
\end{figure*}

\begin{figure*}
\plottwo{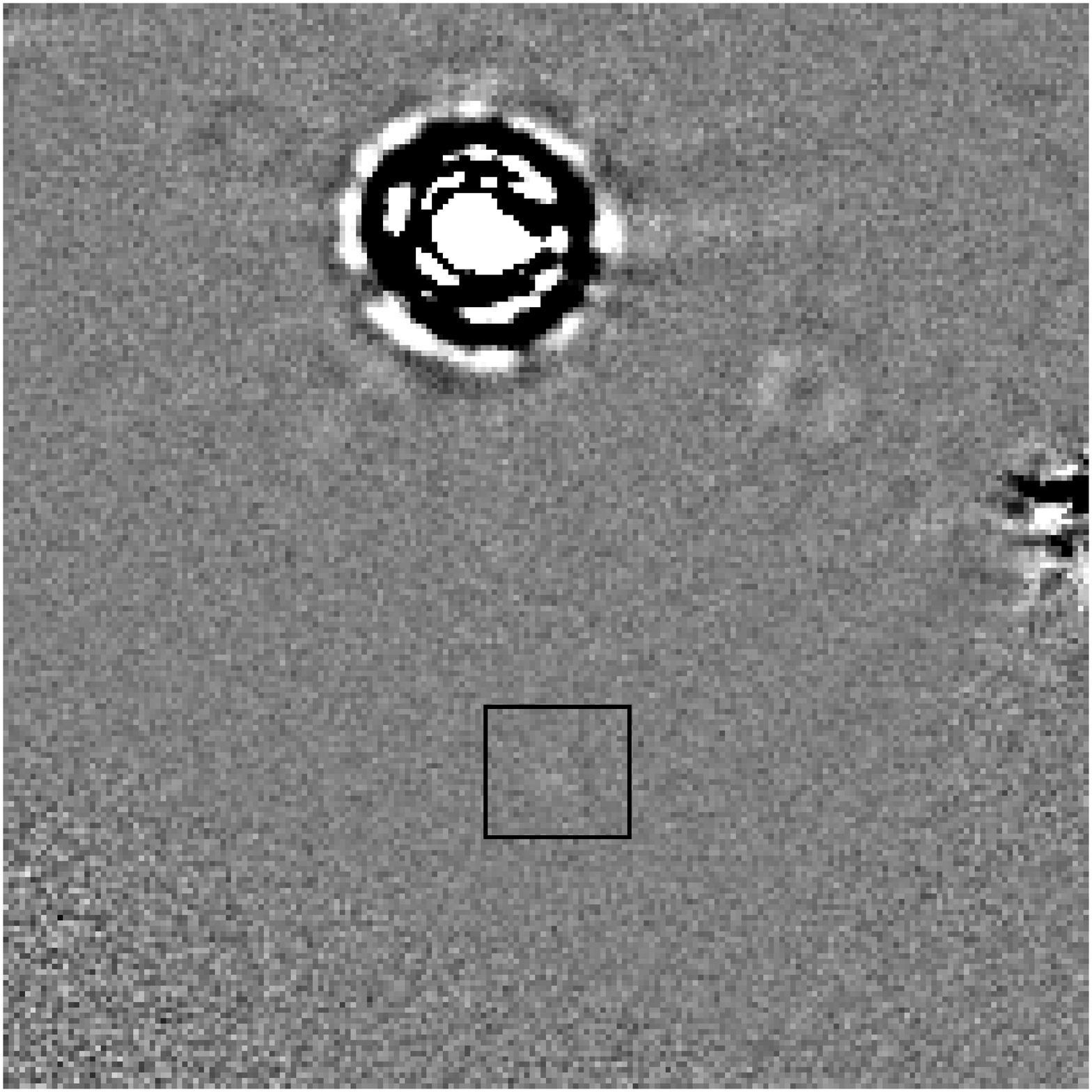}{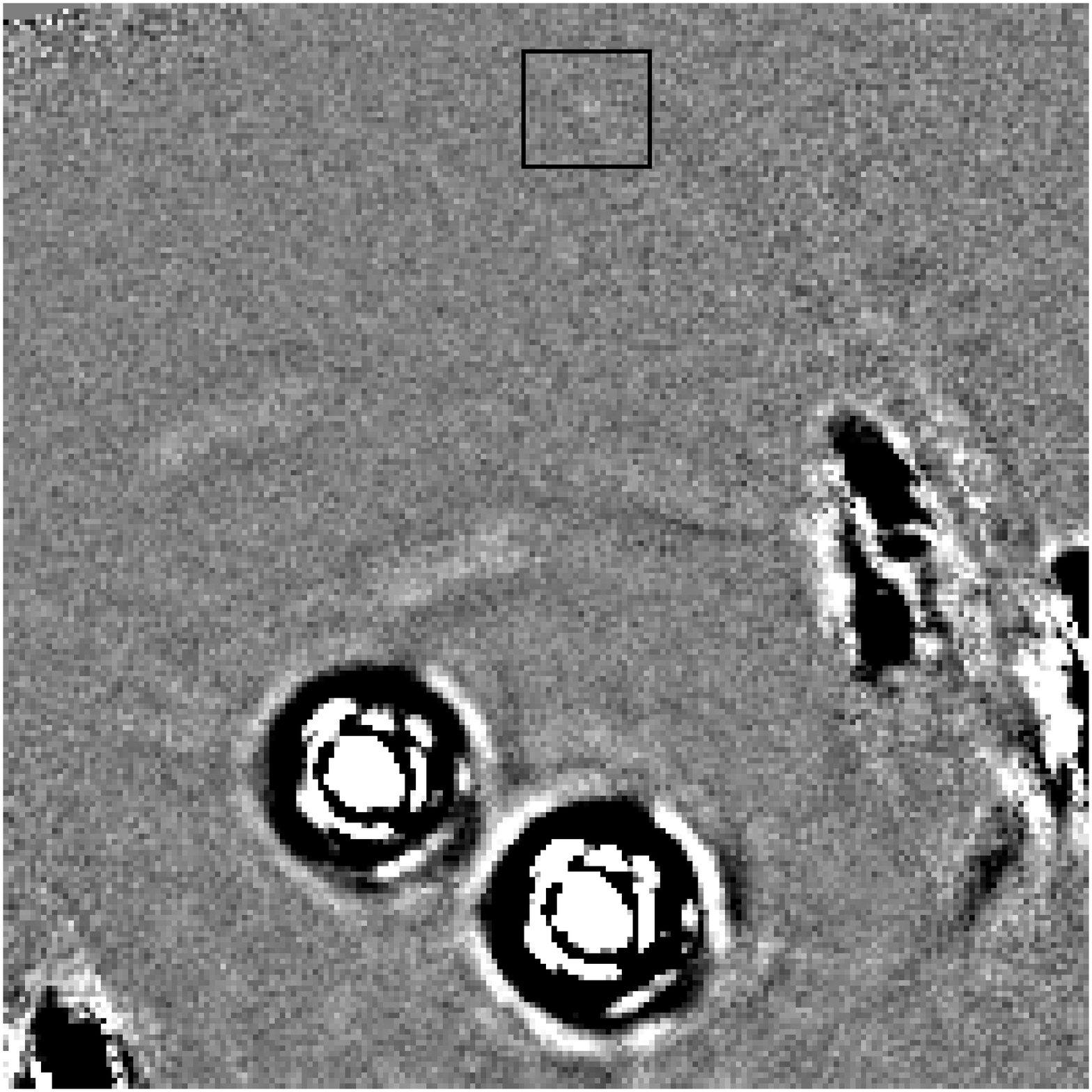}
\caption[Moderate-Separation Companions Detected]{Left,
$L'$ image of GJ 354.1 A, showing the faint background
star we detected.  Right, $L'$ image of binary star
GJ 860, again showing a faint background star.  Each
image is 9.71 arcsec square, contrast stretched the same as
the lower panels in Figure \ref{fig:comp1} to reveal
the faint objects.}
\label{fig:comp2}
\end{figure*}

\begin{figure*}
\plotone{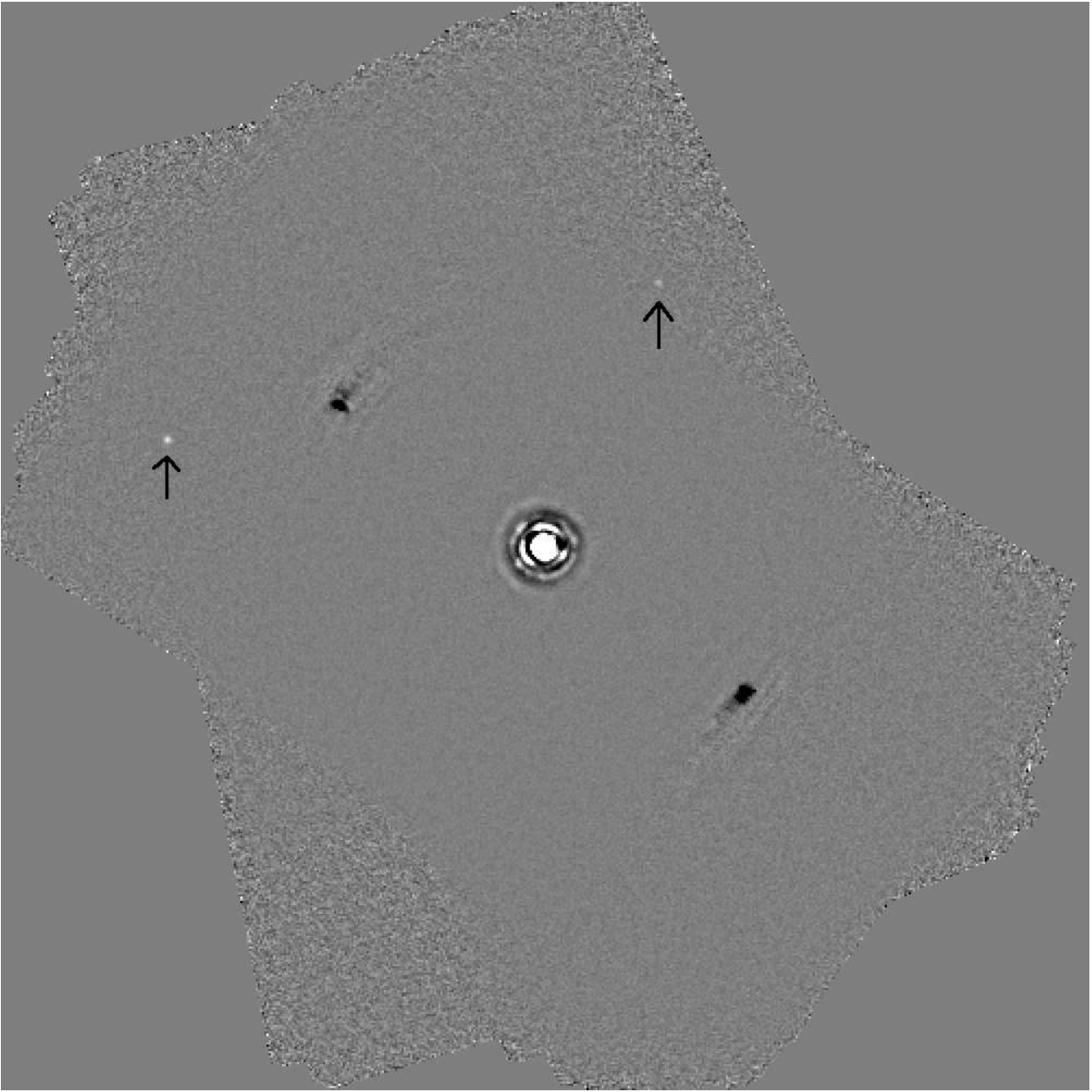}
\caption[Distant Companions to BD+20 1790]{$L'$ image
of BD+20 1790, showing two faint background stars.
Image is 24.29 arcsec square, contrast stretched 3$\times$
less than the images in Figure \ref{fig:comp2}, to
give a clear view of these somewhat brighter stars.}
\label{fig:comp3}
\end{figure*}

\begin{figure*}
\plotone{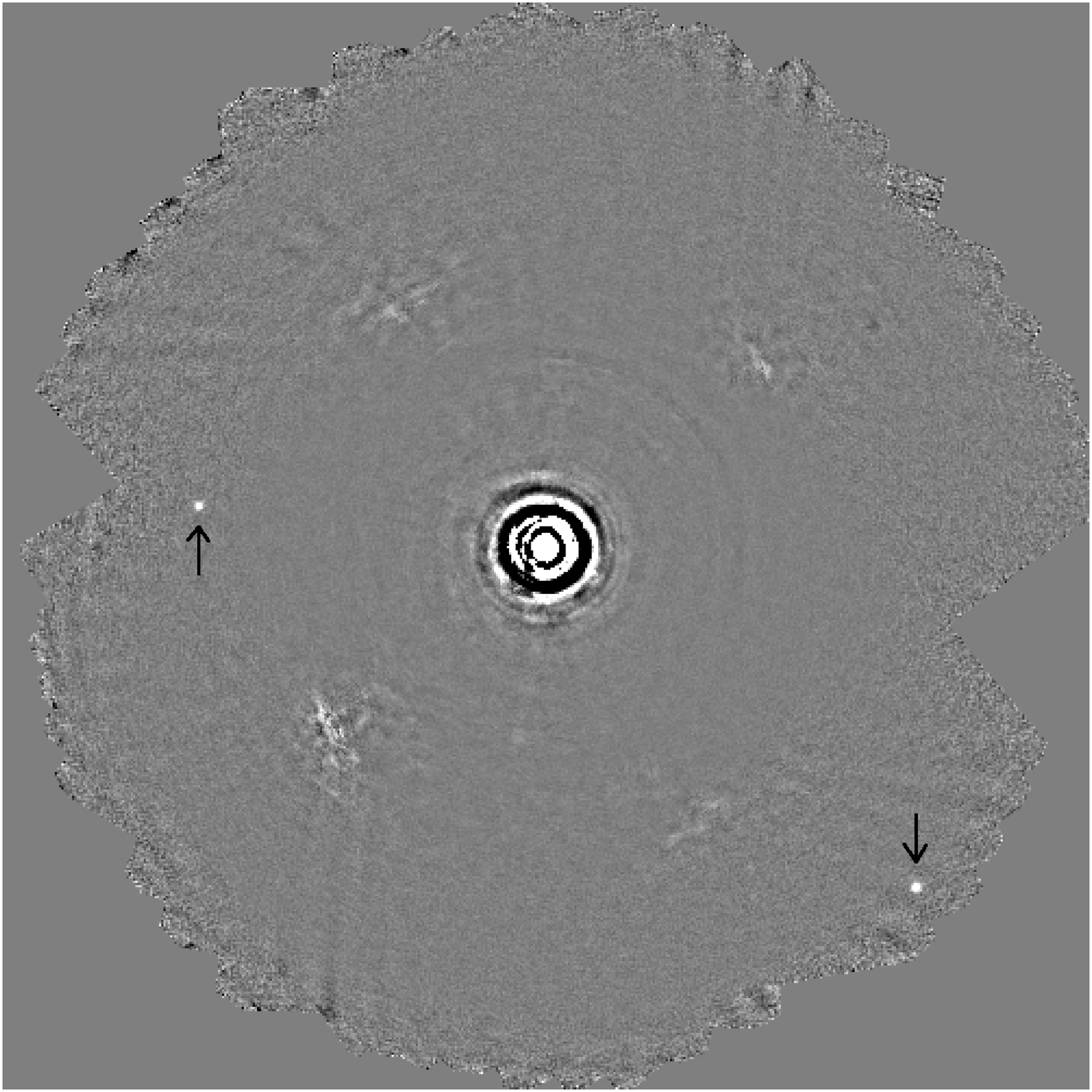}
\caption[Distant Companions to 61 Cyg A]{$L'$ image
of 61 Cyg A, showing two faint background stars.
Image is 24.29 arcsec square, contrast stretched the same
as the previous figure.}
\label{fig:comp4}
\end{figure*}

\begin{figure*}
\plotone{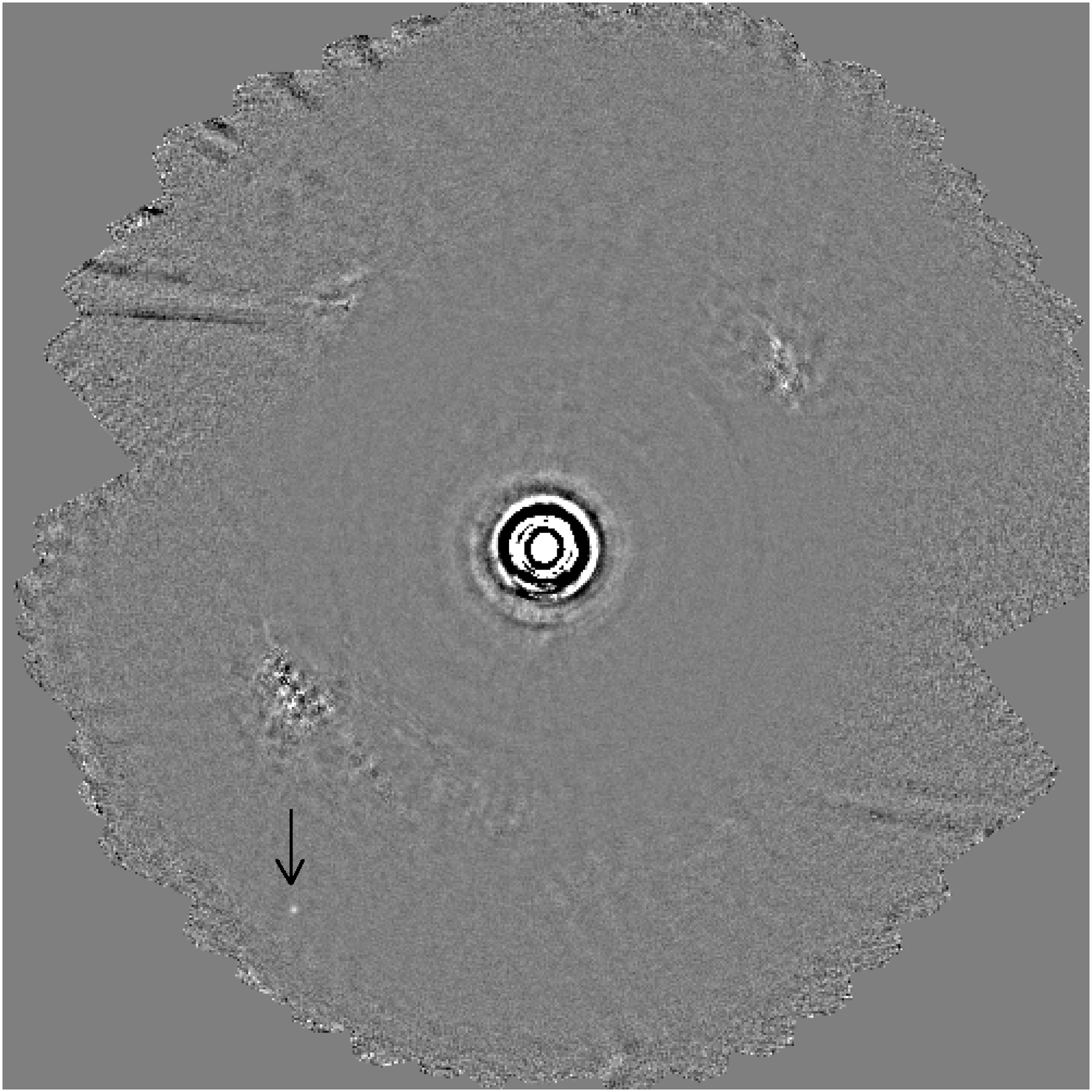}
\caption[Distant Companion to 61 Cyg B]{$L'$ image
of 61 Cyg B, showing a faint background star.
Image is 24.29 arcsec square, contrast stretched the same
as the previous figure.}
\label{fig:comp5}
\end{figure*}

\begin{figure*}
\plotone{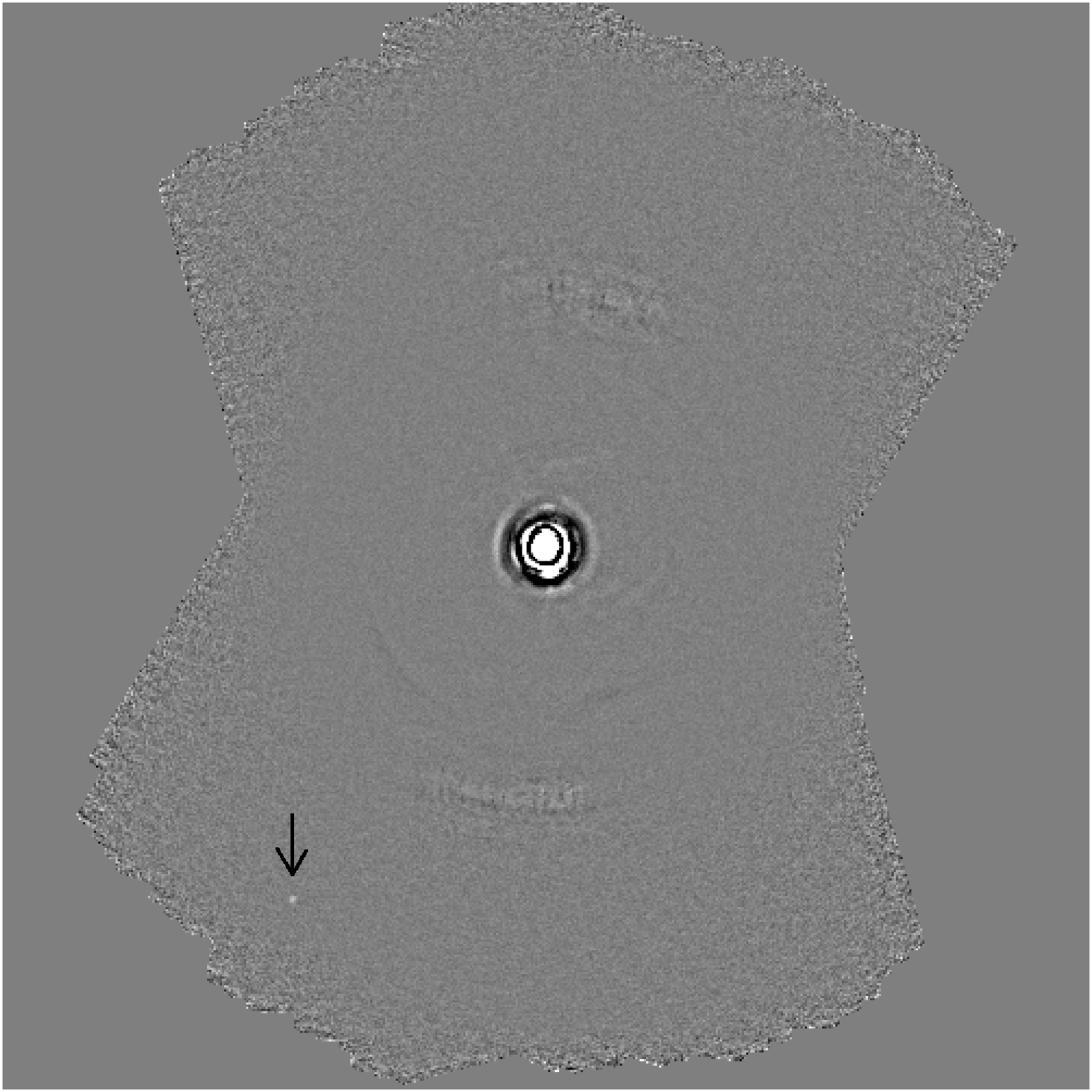}
\caption[Distant Companion to GJ 3860]{$L'$ image
of GJ 3860, showing a faint background star.
Image is 24.29 arcsec square, contrast stretched the same
as the previous figure.}
\label{fig:comp6}
\end{figure*}

Figures \ref{fig:comp1} through \ref{fig:comp6} show all of our
detected companions, except the companion of HD 96064, which
has already been shown in Figure \ref{fig:ymethod}.  Each of
these images is from a `d' method reduction of long exposure
science data. 

\subsection{The Low-Mass Star GJ 3876 B}\label{sec:discovery}
The single discovery of our survey is the low-mass stellar
companion of GJ 3876.  We first detected it on $L'$ images
from April 13, 2006, and confirmed it as a common proper
motion companion in $L'$, $M$, and $K_S$ images taken
on April 11, 2007.  Table 11 gives our
photometric and astrometric results, complete with what
the object's position should have been in April 2007 if
it were a background star.

\begin{deluxetable}{lllccc}
\tablewidth{0pc}
\tablecolumns{6}
\tablecaption{Discovery Data for GJ 3876 B}\label{tab:GJ3876}
\tablehead{\colhead{Date} & \colhead{Sep} & \colhead{PA} & & & \\
\colhead{(yyyy/mm/dd)} & \colhead{(arcsec)} & \colhead{(degrees)} & \colhead{$K_S$} & \colhead{$L'$} & \colhead{$M$}}
\startdata
2006/04/13 & $1.8518\pm0.0038$ & $118.57\pm0.19$ & \nodata &  $10.88\pm0.06$ & \nodata \\
2007/04/11 & $1.8603\pm0.0082$ & $118.64\pm0.24$ & $11.51\pm0.22$ &
$10.79\pm0.08$ & $10.91\pm0.28$ \\
Background & 1.6487 & 113.73 & \nodata & \nodata & \nodata \\
\enddata
\tablecomments{Astrometry and photometry of the single
discovery of our survey, GJ 3876 B.  The first two rows
give actual measured values; the last gives the predicted
position for 2007/04/11 if the object were a background star,
based on the 2006/04/13 position and a proper motion
measurement from \citet{hip}.  The background star hypothesis
is rejected with great confidence.}
\end{deluxetable}

GJ 3876 B is clearly a common proper motion companion.
The distance to the primary star is about 43 pc, based
on the parallax from \citet{hip}.  This translates to
a projected separation of about 80 AU, which suggests
an orbital period of around 700 yr for a one solar
mass primary.  The constant position angle over a year
seems inconsistent with a face-on orbit at this
period, while the formally insignificant increase in
separation may hint at motion in a more inclined
orbit -- however, much more data is needed.

Again using the \citet{hip} distance, the $K_S$ absolute
magnitude of GJ 3876 B is $8.33\pm0.22$.  Based on
the models of \citet{lowmass}, this translates into
a mass of about $0.15\pm0.01$M$_{\sun}$.  This
estimate could be further investigated using our
$L'$ and $M$ band magnitudes, but model magnitudes
for low mass stars in these bands are not readily
available in the literature, and integrating them
from theoretical spectra is beyond the scope of
this paper.

\subsection{Astrometry of Known Bright Binaries}\label{sec:bin}

We calibrated the plate scale and orientation of the
Clio camera using observations of known wide, very long-period
binary stars.  We had previously obtained precise astrometry
of these stars in the optical using the University of 
Arizona's 61 inch Kuiper telescope on Mt. Bigelow
(\citet{binaries}; for more complete data see 
\url{http://www.hopewriter.com/Astronomyfiles/AstrometryPoster.html}).

When selecting our survey sample, we rejected some binary stars
with orbital properties that seemed likely to destabilize any
planets we could detect.  After these rejections, twenty
stars in known binary systems remained in our sample.
Since our AO images allow very accurate astrometry, which
might be useful for refining the orbital parameters of these
nearby binaries, we present our measurements of them in Table \ref{tab:binaries}.

Note that these binary stars change position relatively
quickly, and should not be used for calibration except with a
precise orbital solution.  Those referred to in \citet{binaries}
and the associated website are better for calibration purposes,
but some of them may still have moved significantly since our
measurements.

The measurements in Table \ref{tab:binaries} are averages of
astrometry based on individual frames.  In many cases we had
short, unsaturated images available in addition to our longer,
saturated exposures for planet detection.  This allowed us to
compare the internal precision of both saturated and unsaturated
images, and choose as our final astrometric result the average
of whichever of the two image sets had the smaller internal scatter.
As explained in Section \ref{sec:datproc}, the agreement between
saturated and unsaturated astrometry was generally excellent.
Note that the Table 12 value for our $L'$ observations of
$\xi$ Boo is based on unsaturated images; this was the binary
star with the largest (though still only 0.009 arcsec) saturated/unsaturated
difference in the list given in Section \ref{sec:datproc}.
The uncertainties given in Table \ref{tab:binaries} combine
both measurement scatter and calibration uncertainty.
The latter is generally the larger term, due to the
necessity of calibrating Clio using less precise astrometry
from seeing-limited optical observations.  The true internal scatter
of carefully conducted astrometric observations using Clio and MMTAO is
certainly several times smaller than the uncertainties quoted
in the table. Despite the calibration uncertainties, however, 
clear orbital motion in the star GJ 702 is seen over an 
interval of only ten months. See \citet{binaries} and the 
previously-cited website for an analysis of the challenges
and potential of using AO astrometry for binary star orbital science.  

\begin{center}
\begin{deluxetable}{lcrr}
\tablewidth{0pc}
\tablecolumns{6}
\tablecaption{Astrometry of Binary Survey Targets\label{tab:binaries}}
\tablehead{ & \colhead{Date Obs.} & & \\
\colhead{Star Name} & \colhead{(yyyy/mm/dd)} & \colhead{Sep.(arcsec)} &
\colhead{PA(deg)} }
\startdata
GJ 166 BC & 2006/12/03 & $8.781\pm0.010$ & 153.72 $\pm$ 0.20 \\
HD 77407 AB & 2007/01/05 & $1.698\pm0.004$ & 356.37 $\pm$ 0.20 \\
HD 96064 AB & 2007/01/04 & $11.628\pm0.007$ & 221.61 $\pm$ 0.20 \\
HD 96064 Bab & 2007/01/04 & $0.217\pm0.010$ & 26.60 $\pm$ 4.30 \\
GJ 505 AB & 2006/06/12 & $7.512\pm0.006$ & 104.92 $\pm$ 0.20 \\
$\xi$ Boo AB & 2006/06/10 & $6.345\pm0.006$ & 312.15 $\pm$ 0.20 \\
$\xi$ Boo AB ($M$) & 2006/06/11 & $6.327\pm0.005$ & 312.14 $\pm$ 0.20 \\
GJ 684 AB & 2006/06/11 & $1.344\pm0.004$ & 323.84 $\pm$ 0.20 \\
GJ 702 AB & 2006/06/09 & $5.160\pm0.005$ & 135.79 $\pm$ 0.20 \\
GJ 702 AB ($M$) & 2007/04/11 & $5.290\pm0.004$ & 134.69 $\pm$ 0.20 \\
GJ 860 AB & 2006/06/12 & $2.386\pm0.004$ & 58.55 $\pm$ 0.20 \\
GJ 896 AB & 2006/07/13 & $5.366\pm0.006$ & 86.16 $\pm$ 0.20 \\
HD 220140 AB & 2006/12/03 & $10.828\pm0.007$ & 214.49 $\pm$ 0.20 \\
\enddata
\tablecomments{The internal precision of Clio astrometry 
is considerably better than the uncertainties given here, 
especially for the position angles.  Calibration uncertainty
is important since we had to calibrate the detector from
seeing-limited optical astrometry of wide binaries.
Even so, GJ 702 shows clear orbital motion in the 10 months spanned
by our two measurements.}
\end{deluxetable}
\end{center}

\section{Conclusion} \label{sec:concl}
We have surveyed unusually nearby, mature star systems
for extrasolar planets in the $L'$ and $M$ bands using
the Clio camera with the MMT AO system.  We have
developed a sophisticated image processing pipeline for
data from this camera, including some interesting
innovations.  We have carefully
and rigorously analyzed our sensitivity.  Accurately
determining the sensitivity of AO planet-search images
is a more complex task than, perhaps, has been widely
appreciated.  Our data support the conclusion of \citet{mspeckle} that $5 \sigma$
limits can substantially overestimate the meaningful sensitivity of
an image.  Blind tests involving fake planets inserted in
raw data are the best way to confirm the validity
of any sensitivity estimator, and should be included
in all future planet-search publications.  By extensive
use of such tests, we established a definitive
significance vs. completeness relation for planets in
our data.  This relation is important for use in Monte Carlo simulations
to constrain planet distributions.

We have discovered a physically orbiting $\sim 0.15$M$_{\sun}$ 
binary companion at a projected separation of 80 AU from the star GJ 3876.
We have detected twelve additional candidate faint companions, one of which
is the binary brown dwarf companion of GJ 564 discovered prior to our
observations by \citet{potter}.  The remaining eleven are confirmed to be
background stars.  We note that shorter wavelength surveys, such as that of \citet{GDPS}
in the $H$ band regime, have typically found a much larger number
of background stars, necessitating extensive follow-up observations.
A long wavelength survey such as ours can obtain good sensitivity
to planets, with their very red IR colors, while remaining blind 
to all but the brightest stars.  This reduces the amount of telescope
time spent following up planet candidates that turn out to be background
stars.  

We did not detect any planets, but have set interesting
limits on the masses of planets or other substellar objects
that may exist in the star sytems we surveyed.  In \citet{modeling},
we use extensive Monte Carlo simulations
to show how our null result constrains the mass
and semimajor axis distributions of extrasolar planets
orbiting sun-like stars.

\section{Acknowledgements} 
This research has made use of the SIMBAD online database,
operated at CDS, Strasbourg, France, and
the VizieR online database (see \citet{vizier}).

We have also made extensive use of information and code
from \citet{nrc}. 

We have used digitized images from the Palomar Sky Survey 
(available from \url{http://stdatu.stsci.edu/cgi-bin/dss\_form}),
 which were produced at the Space 
Telescope Science Institute under U.S. Government grant NAG W-2166. 
The images of these surveys are based on photographic data obtained 
using the Oschin Schmidt Telescope on Palomar Mountain and the UK Schmidt Telescope.

We thank the anonymous referee for helpful suggestions.

Facilities: \facility{MMT, SO:Kuiper}

\end{document}